\documentclass[aps,pre,twocolumn,superscriptaddress]{revtex4-2}

\usepackage{lipsum}
\usepackage{amsfonts}
\usepackage{graphicx}
\usepackage{epstopdf}
\usepackage{algorithmic}
\usepackage{bm}
\usepackage{bbold}
\usepackage{amsmath}
\usepackage{braket}
\usepackage{amsopn}
\usepackage{hyperref}
\usepackage{url}
\usepackage{color}
\usepackage{siunitx}
\usepackage[utf8]{inputenc}
\usepackage[T1]{fontenc}
\allowdisplaybreaks[4]
\usepackage{CJK}

\newcommand{\cc}{\mathrm{c}.\mathrm{c}.}

\begin{document}


\title{Dynamics of active nematic defects on the surface of a sphere}




\begin{CJK*}{UTF8}{gbsn}

\author{Yi-Heng Zhang (张一恒)}
\thanks{Corresponding author}
\email[]{zyh@mail.bnu.edu.cn}
\affiliation{Department of Physics, Beijing Normal University, Beijing 100875, China.}

\author{Markus Deserno}
\affiliation{Department of Physics, Carnegie Mellon University, 5000 Forbes Avenue, Pittsburgh, PA 15213, USA}

\author{Zhan-Chun Tu (涂展春)}

\affiliation{Department of Physics, Beijing Normal University, Beijing 100875, China.}

\date{\today}

\begin{abstract}
A nematic liquid crystal confined to the surface of a sphere exhibits topological defects of total charge $+2$ due to the topological constraint. In equilibrium, the nematic field forms four $+1/2$ defects, located at the corners of a regular tetrahedron inscribed within the sphere, since this minimizes the Frank elastic energy. If additionally the individual nematogens exhibit self-driven directional motion, the resulting active system creates large-scale flow that drives it out of equilibrium. In particular, the defects now follow complex dynamic trajectories which, depending on the strength of the active forcing, can be periodic (for weak forcing) or chaotic (for strong forcing). In this paper we derive an effective particle theory for this system, in which the topological defects are the degrees of freedom, whose exact equations of motion we subsequently determine. Numerical solutions of these equations confirm previously observed characteristics of their dynamics and clarify the role played by the time dependence of their global rotation. We also show that Onsager's variational principle offers an exceptionally transparent way to derive these dynamical equations, and we explain the defect mobility at the hydrodynamics level.
\end{abstract}


\maketitle

\end{CJK*}

\section{Introduction}
Topological defects show up in a surprising variety of ordered systems, and whenever they do give rise to fascinating emergent physics. They have been observed in the study of liquid crystals  \cite{Chandrasekhar1986,Kurik1988,deGennes1993,Kikuchi2002}, crystalline solids  \cite{Taylor1934,Friedel1964,Nabarro1967}, cell assemblies  \cite{Duclos2016,Saw2017,Kawaguchi2017}, superfluid vortices  \cite{Osborne1950,Wilks1987,Vollhardt1990}, magnetic skyrmions  \cite{Bogdanov1989,Bogdanov1994,Bogdanov2006,Binz2006,Tewari2006,Binz2009,Neubauer2009,Pappas2009,Nagaosa2013}, and cosmology  \cite{Turok1989,Pargellis1991,Chuang1991}. Since defects cannot continuously vanish (they typically only pair-create or -annihilate), they constitute long-lived markers of the field that forms them, and hence their dynamics can provide deep insights into the long-scale time evolution of such systems  \cite{Bray1994}. In recent experiments, Keber \emph{et al.}  \cite{Keber2014} have fabricated a spherical nematic by confining kinesin-microtubule bundles onto the surface of a spherical lipid vesicle, and adding ATP to the kinesin motors renders this system active, \emph{i.e.}, self-driven. The ATP-induced activity can drive the system far from a static equilibrium state, leading to a novel defect dynamics in which the active flow competes with dissipative relaxation processes, as well as elastic forces that arise from the nematic field itself and the curvature of the substrate. Significant theoretical understanding has been gained how such active driving forces affect the defect dynamics in a \emph{planar} nematic liquid crystals \cite{Giomi2013,Pismen2013,Thampi2013,Thampi2014,Thampi2014V2,Giomi2014,Gao2015,Shankar2018,Liverpool2018}; in contrast, a comprehensive theoretical framework for the topologically distinct case of spherical confinement lags noticeably behind.

The motion of defects can be regarded as a particular way in which the active nematic field collectively manifests itself: the field moves, and the defects ``ride'' with it. Hence, in order to understand this motion, it is natural to start from the traditional nematic hydrodynamic equations \cite{ericksen1959,ericksen1961,leslie1966,leslie1968,Beris1994}, amended by a suitably chosen active stress  \cite{Simha2002,Marchetti2013,Prost2015,Ramaswamy2017,Doostmohammadi2018}. In this way, one arrives at a continuum theory that describes the evolution of the nematic order and its underlying flow velocity, and one may thereby predict how the defects are transported along. We also emphasize that the curvature of the substrate can play an important role in the active nematic hydrodynamics, not only the intrinsic curvature of a sphere, but also the extrinsic curvature, like the curvature of a cylindrical surface \cite{Napoli2016,Napoli2020}. While conceptually fairly clean, the resulting partial differential equations do not afford illuminating analytical solutions, and therefore a substantial amount of work has been dedicated to numerically solving them, which has indeed provided much novel insight into the collective behavior of spherical active nematic systems \cite{Shin2008,Rui2016,Henkes2018}. 

However, the topological nature of this problem strongly suggests an alternative viewpoint. Recall that defects cannot continuously vanish or arise. Their topological discreteness ``protects'' their existence, or, in physics parlance, it equips them with a \emph{conservation law}. More precisely: since a defect's topological index can  only in finite steps (say, integral, or half-integral), defects require other defects to change their overall number---meaning that they typically create and annihilate in pairs. This, of course, equips them with a property we commonly associate with elementary \emph{particles} (such as electrons), which can be created or destroyed only in particle-antiparticle pairs. Their longevity suggests that it should be expedient to formulate an \emph{effective} theory that conceives of these discrete particles as the essential degrees of freedom, which in turn interact by force fields that emerge as ``leftovers'' of the originally strongly coupled field. In our context, this perspective suggests that we view topological defects in a nematic as particles, whose dynamics is given by the force balance between the effective friction and the elastic interaction. Such an effective theory (in soft matter language more usually called ``coarse-grained'') has indeed been successfully used to describe the defect dynamics in planar nematics \cite{DAFERMOS1970,IMURA1973,Pismen1990,Ryskin1991,lubensky1992,Denniston1996,Park1996,Nelson2002,Sven2002,kats2002,Denniston2002,sonnet2005,Vitelli2006,Sonnet2009,Radzihovsky2015}. Since the Euler characteristic of a (simply connected) plane is zero, the total sum of all indices of defects is zero, so that the number of defects equals that of the ``antidefects'' (meaning that there are exactly as many $+1/2$ defects as $-1/2$ defects). Hence, the characteristic physical processes are the creation and annihilation of defects in pairs out of and into an otherwise structureless (``trivial'') vacuum ground state---a scenario that is indeed excellently described by the effective particle approach \cite{Giomi2013,Pismen2013,Giomi2014,Rui2018,Liverpool2018}. 

Keber \emph{et al.} have taken the first steps and developed a minimal model of defects as effective particles moving on a spherical surface  \cite{Keber2014}. Following their lead, various important studies of the dynamics of active defects in curved surfaces have been proposed. In a recent publication Khoromskaia and Alexander \cite{Khoromskaia2017} refined the Keber model by calculating the active flow via the Stokes equation of the active nematic. Although their theoretical model explained some of the experimentally observed phenomena, it does not yet provide a detailed connection between the coarse-grained dynamics of defects and the active nematic hydrodynamic equations. Furthermore, nematic defects have a broken rotational symmetry, and hence their effective description must include not just their \emph{position} as the sole degree of freedom, but also a vectorial \emph{orientation}  \cite{Keber2014,Vromans2016,Tang2017}. This is especially obvious for the $+1/2$ defects, which self-propel in a well-defined direction. Moreover, broken rotational symmetry implies that these oriented particles are not just subject to effective isotropic forces; rather, forces will depend on mutual orientation, and there will also be effective \emph{torques}---all of which will affect the resulting dynamics \cite{Keber2014,Vromans2016,Tang2017,Shankar2018,shankar2019}. Khoromskaia and Alexander have indeed explored the impact of this orientational degree of freedom on the active behavior of these systems \cite{Khoromskaia2017}, but the resulting dynamics, and especially the role of effective elastic torques, is still not understood well. In a recent paper \cite{Aidan2020}, Brown has accounted for the orientation dynamics of defects on the sphere, and partly explained the experimental and simulated trajectories of defects. More generally, there is significant interest to understand how confinement impacts active nematodynamics in cases other than the surface of a sphere, especially for different topologies and nontrivial Gaussian curvature, such as a torus, and some important progress along those lines has been made \cite{Bowick2004,ellis2018,Pearce2019}.

In the present paper, we aim to develop a more detailed effective field theory for defects in active nematics that are confined to the surface of a sphere, based on active nematic hydrodynamic. The main philosophy is no different from the planar case, and hence we expect the essential scale separation to work just as well. However, there is a very significant difference that renders this case more difficult, but also more interesting: the Euler characteristic of a spherical surface is $+2$, and so we have broken ``matter-antimatter symmetry''. In other words, \emph{the vacuum is not empty}. Moreover, since it is well known that in an active nematic the $+1/2$ defects are self propelled \cite{Giomi2014}, this also implies that \emph{the vacuum is never at rest}. Hence, particle pairs are not created into an otherwise quiescent vacuum but are immediately advected with the preexisting ground-state flow---which we hence need to understand first. We follow the ideas proposed by Tang and Selinger  \cite{Tang2019} to advance the theory of a dry active nematic liquid crystal confined onto a spherical substrate. Specifically, we show how to make use of an elegant variational principle due to Onsager \cite{Vertogen1983,Vertogen1989,Sonnet2001,Doi2011,Sonnet2012,Tang2019} in order to transition from the infinite-dimensional nematic field theory to the finite-dimensional effective field theory of oriented defects. As a particular result, we derive the \emph{anisotropic mobility coefficient matrix} of defects. We then show that our effective theory fully reproduces the complex periodic trajectory of the four $+1/2$ defects, and how it depends on system size and the strength of the active forcing, as reported in earlier numerical work and experiments \cite{Keber2014,Rui2016,Henkes2018}. Our theory includes the full dynamics of a defect's orientation, which permits us to identify its importance for the resulting particle motions.

We have organized the content of our paper as follows. In Sec. II, we outline the hydrodynamic model we employ in our discussion, followed by the theoretical approach we use to separate the dynamical variables of the defects from the temporal evolution of the active nematic field. In Sec. III, we look at some of the predictions of our theory, including the well-known periodic oscillation of the defect configuration under weak forcing, and the chaotic trajectories under stronger driving conditions. Sec. IV summarizes our conclusions and lists a number of limitations of our theory.

\section{MODELS AND METHODS}
\subsection{Minimal model}

The nematic liquid crystal is locally characterized by two variables: order and flow. We can describe the order by the nematic tensor
\begin{equation}
    Q^{ab}=q\left(\mathrm{n}^a\mathrm{n}^b-\frac12g^{ab}\right) \ ,
\end{equation}
in which $q$ is the magnitude of the nematic tensor representing its average alignment in a small region, and $g^{ab}$ is the (inverse) metric tensor;
the unit vector $\mathbf{n}=\mathrm{n}^a\mathbf{e}_a$ denotes the local nematic director, whose components
$\mathrm{n}^a=(\cos{\psi}, \sin{\psi})$ refer to the spherical orthonormal basis $\{\mathbf{e}_\theta,\mathbf{e}_\phi\}$ \cite{deGennes1993} and are fully specified by a single scalar function $\psi(\theta,\phi)$ of position. (As usual, a repeated upper and lower index implies a tensor contraction.) The (incompressible) flow is characterized by the velocity field $V^a$. These two variables $\{Q^{ab},V^a\}$ satisfy the hydrodynamic equations of the active nematic, with a constant density $\rho$ on the curved surface \cite{Edwards1990,Olmsted1992,Beris1994,Olmsted1997,Olmsted1999,Pearce2019}:
\begin{subequations}
\begin{align}
    \mathring{Q}^{ab} &= S^{ab} -\frac{1}{\gamma}\frac{\delta F_{\mathrm{LdG}}}{\delta Q^{ab}} \ , \label{BEE}\\
    \rho\mathring{V}^a &= \alpha\nabla_bQ^{ba}+\nabla_b\Pi^{ab}-\zeta V^a \ . \label{NSE}
\end{align}
\end{subequations}
Here, the ring-operator $(\mathring{\;\;})\equiv \mathrm{D}/\mathrm{D}t\equiv\partial/\partial_t+V^c\nabla_c$ denotes the \emph{covariant material derivative}, $\gamma$ is the rotational viscosity, and $F_{\mathrm{LdG}}$ is the Landau-de Gennes free energy of the system \cite{deGennes1993}, which contains the homogeneous phase energy $F_\mathrm{p}$, as has also recently been derived by dimensional reduction from a full three-dimensional Landau-de~Gennes functional \cite{Napoli2012}:
\begin{align}
    F_\mathrm{p}=\int\mathrm{d}S\,\left(\frac{A}{2}Q_{ab}Q^{ab}+\frac{B}{4}\left(Q_{ab}Q^{ab}\right)^2\right) \ ; \label{phener}
\end{align}
and, by assuming (for simplicity) the one-con\-stant-ap\-prox\-i\-ma\-tion, the Frank energy $F_\mathrm{e}$ caused by spatial distortions:
\begin{align}
    F_\mathrm{e}=\frac{K}{4}\int\mathrm{d}S\,(\nabla_cQ^{ab})(\nabla^cQ_{ab}) \ . \label{dfra}
\end{align}
The tensor $S^{ab}$ in Eq. (\ref{BEE}) represents the coupling between the director field, the (symmetric) strain rate tensor $A_{ab}=\nabla_{[a}V_{b]}$ and the (antisymmetric) vorticity $\omega_{ab}=\nabla_{(a}V_{b)}$, with the parameter $\xi$ reflecting the flow aligning of the nematic field \cite{Olmsted1992,Kruse2004,Marenduzzo2007,Liverpool2008,Lau2009,Giomi2012}:
\begin{equation}
    S^{ab} = \xi qA^{ab}+Q^{ac}\omega_c{}^b-\omega^a{}_cQ^{cb}.\label{sab}
\end{equation}
The activity enters into the system through the active stress $\alpha\nabla_bQ^{ba}$ in the covariant Navier-Stokes equation (\ref{NSE}), where the parameter $\alpha$ controls the strength of the activity. The system is termed ``contractile'' if $\alpha>0$, and ``extensile'' if $\alpha<0$ \cite{Ramaswamy2010,Giomi2012,Marchetti2013}. The passive stress is denoted by $\Pi^{ab}$. The substrate friction $\zeta V^a$ arises from the damping force between the active nematic and the spherical substrate beneath it. If we neglect the inertia term in Eq. (\ref{NSE}) and assume the substrate friction is much larger than the other remaining dissipation terms, the only contribution to the velocity field will be the active flow \cite{Shankar2018,shankar2019}:
\begin{align}
    V^a=\frac{\alpha}{\zeta}\nabla_bQ^{ba} \ . \label{acfl}
\end{align}

\subsection{Coarse grained dynamics of +1/2 defects}

\begin{figure}
\includegraphics[width=\linewidth]{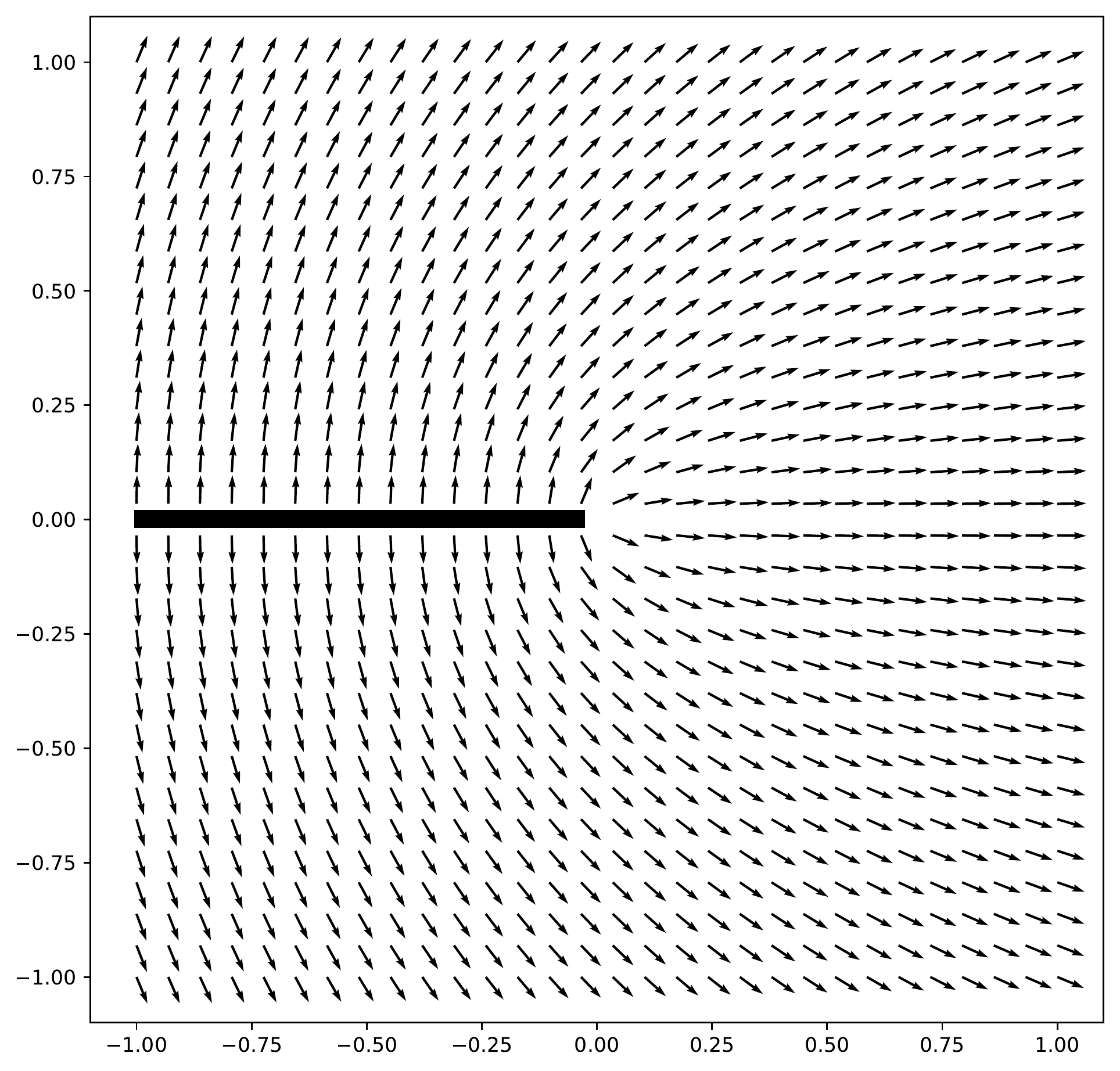}
\caption{Vector field with a $+1/2$ defect at the center and a discontinuity line along the negative real axis due to the ``phase shift'' of $\pi$.\label{defect}}
\end{figure}

In our particular (two-dimensional) case, the index or charge of a vector field's defect may be defined as follows: take a closed path around an isolated defect and follow the orientation of the vector field as you traverse that path once, then the index is the number $s$ of times which this vector field rotates before you arrive back at the starting point (meaning that it acquires a ``phase shift'' of $2\pi s$ along this loop). We give a schematic of a $+1/2$ defect in Fig.~\ref{defect}. It is clear to see that if we start from any point at the negative real axis and trace the vector direction around the original point clockwise, the vector acquires a ``phase shift'' of $\pi$. The continuum model of nematic liquid crystals on a spherical surface can be locally described by a vector field on a spherical surface \cite{deGennes1993}. Because the Euler characteristic of a spherical surface is $+2$, the existence of defects is inevitable: summing the indices of all defects must yield $+2$. The only difference is that the apolar nature of a liquid crystal's \emph{director field} (meaning, it is already invariant under a \ang{180} in-plane rotation) permits half-integral indices for the defect points. In the present work we will restrict our discussion to the case of four $+1/2$ defects, which is the ground state of the nematic director field on the spherical surface.

The motion of defects is a collective effect of the nematic field, so the dynamic information of the defects is contained in the hydrodynamic equations we have just given. In order to extract the variables we care about, which are the position and the orientation of each defect, and obtain an effective field theory of the defects, we will now introduce the framework of Onsager's variational principle as an alternative description of the active nematic hydrodynamics \cite{Vertogen1983,Vertogen1989,Sonnet2001,Doi2011,Sonnet2012,Napoli2016}.

The so-called \emph{Rayleighian} corresponding to the Beris-Edwards equation (\ref{BEE}) is \cite{Stewart2004}:
\begin{align}
    \mathfrak{R}=\frac{\mathrm{d}}{\mathrm{d}t}F_{\mathrm{LdG}}+\frac{\gamma}{2}\int\mathrm{d}S\left(\mathring{Q}^{ab}-S^{ab}\right)\left(\mathring{Q}_{ab}-S_{ab}\right).\label{grel}
\end{align}
Its major use lies in the fact that the Beris-Edwards equation follows from minimizing the Rayleighian (\ref{grel}) with respect to $\partial_tQ^{ab}$.

\begin{figure}
\includegraphics[scale=0.7]{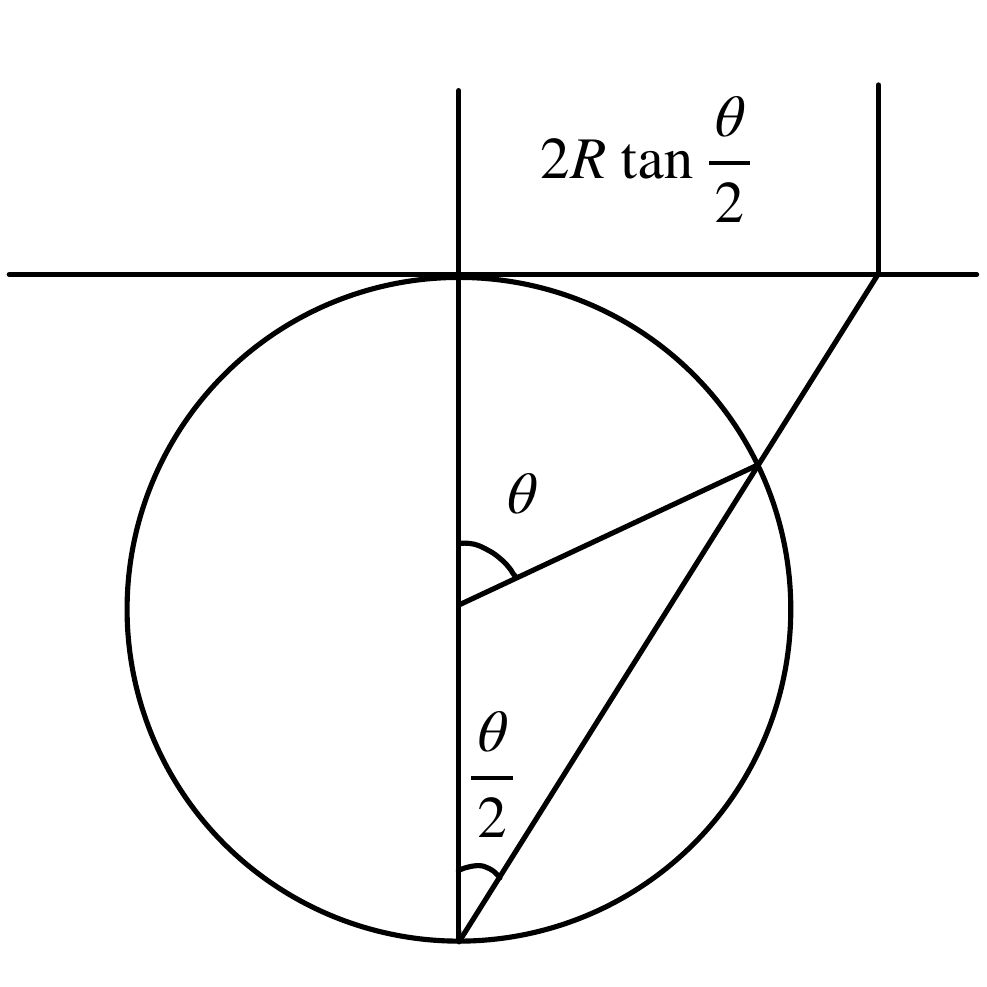}
\caption{The stereographic mapping of the spherical surface onto the complex plane.\label{representation}}
\end{figure}

Now, if we assume that the timescale of the active flow is much slower than the characteristic timescale by which the director field $\mathbf{n}$ relaxes, it is reasonable to assume that $\mathbf{n}$ stays close to its equilibrium configuration during its active motion \cite{Giomi2014,Khoromskaia2017}. By using Riemann's stereographic projection $z(\theta, \phi)=2R\tan{\frac{\theta}{2}}\mathrm{e}^{\mathrm{i}\phi}$, which maps the spherical coordinate $(\theta,\phi)$ of the sphere with radius $R$ to the complex plane (such that the north pole becomes the origin and the south pole maps to complex infinity; see Fig.~\ref{representation}), the equilibrium configuration of the director field with four $+1/2$ topological defects at specified positions was given by \cite{lubensky1992,Khoromskaia2017}
\begin{align}
    \psi(z)
    &=\Omega-\phi+\frac12\sum_{k=1}^4{\psi_k}\nonumber\\
    &=\Omega-\phi+\frac12\sum_{k=1}^4{\mathfrak{Im}\ln(z-z_k)} \ , \label{cubs}
\end{align}
in which $\mathfrak{Im}$ picks out the imaginary part of a complex number, $z_k=2R\tan{\frac{\theta_k}{2}}\mathrm{e}^{\mathrm{i}\phi_k}$ is the position of the $k$-th defect, and $\psi_k$ represents the director field excited by a single $+1/2$ defect at $z_k$.
This is nothing but the solution of the Euler-Lagrange equation of the Frank energy, efficiently expressed in complex notation. For the purpose of simplicity, we will take the \emph{equilibrium} solution (\ref{cubs}) as the configuration of the \emph{dynamic} director field, thus omitting higher order corrections from the velocity of defects. This is in accord with our assumption of a small active flow.

If we look at the director field near the $j$-th defect, and use $\Psi_j$ to denote the angle between the symmetry axis of the defect and $\mathbf{e}_\theta$, then
\begin{align}
    \Psi_j=2\Omega-\phi_j+\sum_{k\neq j}\arg \left(z_j-z_k\right) \ .\label{angle}
\end{align}
According to Refs.~\cite{Vromans2016,Tang2017}, this expression of $\Psi_j$ represents the optimal relative orientation of defects. It implies an extra constraint for the time evolution of the orientation of defects. We will illustrate later that this constraint is consistent with our assumption of the timescale separation. This also shows that $\Omega$ describes a global rotation of all the defects; we will demonstrate in the next section that it has a very important effect on their dynamics.

As a consequence of the small active flow assumption, the scalar order $q$ varies only weakly outside the defects. We hence assume that $q$ is constant away from the defect but melts to zero inside a small core region, because the director would have to assume every orientation at the center. We will not try to describe the precise way in which the order vanishes towards the center, and it will indeed not be important. Instead, we will assume that any integral area we are concerned about must exclude the defect core, and that outside the defects $q$ is constant (we will assume $q=0.62$ in our subsequent numerical examples to permit easier comparison with Ref.~\cite{Rui2016}). And since $q$ is now constant over the entire domain of interest, the phase energy $F_\mathrm{p}$ does not contribute to the dynamics. We can therefore restrict the free energy $F_{\mathrm{LdG}}$ to the Frank energy contribution $F_{\mathrm{e}}$.

With this ansatz, the time evolution of the director field is identical to the evolution of the position and the direction of defects, which means
\begin{align}
    \partial_t\psi
    &=\dot\Omega+\dot\theta_k\frac{\partial\psi_k}{\partial\theta_k}+\dot\phi_k\frac{\partial\psi_k}{\partial\phi_k}\nonumber\\
    &=\dot\Omega+\frac12\sum_{k=1}^4 \Big(\frac{\mathrm{i}}{z-z_k}\dot z_k+\cc\Big)
\end{align}
and the dot $(\dot{\;\;})\equiv\mathrm{d}/\mathrm{d}t$ means a total derivative with respect to time. Now, according to Onsager's variational principle, the minimization of the Rayleighian (\ref{grel}) with respect to the dynamic variables of concern (here: velocities of positions and global orientations of defects), will directly give us the dynamics of the defects. Hence, their equations of motion can be succinctly written as
\begin{subequations}
\begin{align}
    \frac{\partial}{\partial\dot{\theta}_k}\mathfrak{R}&=0\label{vot} \ , \\
    \frac{\partial}{\partial\dot{\phi}_k}\mathfrak{R}&=0\label{vop} \ , \\
    \frac{\partial}{\partial\dot{\Omega}}\mathfrak{R}&=0\label{voo} \ .
\end{align}
\end{subequations}
By doing this, the infinite numbers of degrees of freedom in the nematic field are now reduced to nine degree of freedom for the four defects, which is our main aim in this work. 

Now we can rewrite the Rayleighian (\ref{grel}) as

\begin{align}
    \mathfrak{R}&=\frac{\mathrm{d}}{\mathrm{d}t}F_{\mathrm{e}}+\gamma q^2\int\mathrm{d}S\left(\partial_t\psi\right)^2\nonumber\\
    &\phantom{=}\;-2\frac{\gamma\alpha q^2}{\zeta}\int\mathrm{d}S\nabla_b\left(\nabla_a\psi-A_a\right)Q^{ab}\partial_t\psi \ , \label{disp}
\end{align}
in which $A_a=\mathbf{e}_\theta\cdot\partial_a\mathbf{e}_\phi$ is the spin connection. Details of the derivation are presented in Appendix A.

The Frank energy, in turn, is given by \cite{lubensky1992,Park1996,Nelson2002,Vitelli2006}
\begin{align}
    F_{\mathrm{e}}=-\frac{\pi K q^2}{8}\sum_{j\neq k}\ln \left(1-\cos\beta_{jk}\right) + \mathrm{const.} \ ,
\end{align}
where $\beta_{jk}$ is the angular distance between defects $j$ and $k$, whose cosine can be expressed via
\begin{align}
    \cos\beta_{jk}=\cos\theta_j\cos\theta_k+\sin\theta_j\sin\theta_k\cos(\phi_j-\phi_k) \ .
\end{align}
After making use of these, we can now write the dynamic equations (\ref{vot}--\ref{voo}) as
\begin{subequations}
\begin{align}
    M_{jk}\dot\theta^k+N_{jk}\dot\phi^k+\Theta_{j}\dot\Omega-T_{j}&=
    -\frac{1}{2q^2\gamma}\partial_{\theta_j}F_{\mathrm{e}} \ , \label{D1}\\
    N_{kj}\dot\theta^k+P_{jk}\dot\phi^k+\Phi_{j}\dot\Omega-S_{j}
    &=
    -\frac{1}{2q^2\gamma}\partial_{\phi_j}F_{\mathrm{e}} \ , \label{D2}\\[0.5em]
    \Theta_{k}\dot\theta^k+\Phi_{k}\dot\phi^k+4\pi R^2\dot\Omega-L
    &=
    0 \ , \label{D3}
\end{align}
\end{subequations}
where we introduced the following abbreviations for the eight different integrals that emerge in the process:
\begin{subequations}
\begin{align}
    M_{jk}&=\int\mathrm{d}S\,\partial_{\theta_j}\psi_j\partial_{\theta_k}\psi_k \ , \label{M_jk}\\
    N_{jk}&=\int\mathrm{d}S\,\partial_{\theta_j}\psi_j\partial_{\phi_k}\psi_k \ , \label{N_jk}\\
    P_{jk}&=\int\mathrm{d}S\,\partial_{\phi_j}\psi_j\partial_{\phi_k}\psi_k \ , \label{P_jk}\\
    \Theta_{j}&=\int\mathrm{d}S\,\partial_{\theta_j}\psi_j \ , \label{Theta_j}\\
    \Phi_{j}&=\int\mathrm{d}S\,\partial_{\phi_j}\psi_j \ , \label{Phi_j}\\
    L&=\frac{\alpha}{\zeta}\int\mathrm{d}S\,\nabla_b\left(\nabla_a\psi-A_a\right)Q^{ab} \ , \label{L}\\
    T_{j}&=\frac{\alpha}{\zeta}\int\mathrm{d}S\,\nabla_b\left(\nabla_a\psi-A_a\right)Q^{ab}\partial_{\theta_j}\psi_j \ , \label{T_j}\\
    S_{j}&=\frac{\alpha}{\zeta}\int\mathrm{d}S\,\nabla_b\left(\nabla_a\psi-A_a\right)Q^{ab}\partial_{\phi_j}\psi_j \ . \label{S_j}
\end{align}
\end{subequations}
Notice that in all cases the integral area excludes a defect core of size $r_k=r(1+|z_k|^2/4R^2)$, which is the projection image of the $k$-th defect core radius $r$ at the complex plane \cite{lubensky1992}. Complete analytical expressions for all of these integrals, and some of the technically tedious details for how to obtain them, are given in Appendices B, C, and D.

If we introduce the characteristic timescale $\tau=\gamma R^2/2K$ and define the corresponding dimensionless time $\tilde{t}=t/\tau$, then the dimensionless dynamic equations of the defects of an active nematic on a sphere can be succinctly written as
\begin{subequations}{}
\begin{align}
    m_{jk}\frac{\mathrm{d}\theta^k}{\mathrm{d} \tilde{t}}+n_{jk}\frac{\mathrm{d}\phi^k}{\mathrm{d} \tilde{t}}-t_j&=-\partial_{\theta_j}f \ , \label{SDT}\\
    n_{kj}\frac{\mathrm{d}\theta^k}{\mathrm{d} \tilde{t}}+p_{jk}\frac{\mathrm{d}\phi^k}{\mathrm{d} \tilde{t}}-s_j&=-\partial_{\phi_j}f \ , \label{SDP}\\
    \frac{\mathrm{d}\Omega}{\mathrm{d} \tilde{t}} &= -\frac{\Phi_k}{4\pi R^2}\frac{\mathrm{d}\phi^k}{\mathrm{d} \tilde{t}} \ , \label{SDO}
\end{align}
\end{subequations}
with the corresponding scaled variables
\begin{align}
    &m_{jk}=\frac{M_{jk}}{R^2},\quad n_{jk}=\frac{N_{jk}}{R^2},\quad p_{jk}=\frac{P_{jk}}{R^2}-\frac{\Phi_j\Phi_k}{4\pi R^4},\nonumber\\ 
    &t_j=\frac{\gamma}{2K}T_j,\quad s_j=\frac{\gamma}{2K}S_j,\quad f=\frac{F_{\mathrm{e}}}{4q^2K} \ . \nonumber
\end{align}

\section{RESULTS}

We have established the dynamics of defects in the spherical active nematic and expressed the mobility coefficient matrices (\ref{M_jk}--\ref{P_jk}) in terms of active nematic hydrodynamics parameters. Now we would like to compare our model with previous phenomenological theories and numerical studies.

The analytical expressions (\ref{cT_j}) and (\ref{cS_j}) for the integrals $T_j$ and $S_j$, worked out in Appendix D, suggest that these are actually the components of the mean active back flow around the $j$-th defect---in agreement with the calculations given by Khoromskaia and Alexander \cite{Khoromskaia2017}. We hence notice that the diagonal elements of the mobility matrices in Eq. (\ref{SDT}) and (\ref{SDP}) are
\begin{subequations}
\begin{align}
    m_{jj} &= \frac{\pi}{8}\Big(2\ln\frac{2R}{r}-1\Big) \ ,\label{mjj} \\
    p_{jj} &= \frac{\pi}{8}\Big(2\ln\frac{2R}{r}-1\Big)\sin^2\theta_j \ .\label{pjj}
\end{align}
\end{subequations}
If the radius $R$ of the sphere is very large compared to the core radius $r$ of the defects, and defects do not approach each other too closely during their motion, the nondiagonal elements are negligible. In this situation, Eq. (\ref{SDT}) and (\ref{SDP}) are identical to the approximations proposed in Ref.~\cite{Khoromskaia2017}, and therefore the physical picture is similar here. The motion of defects reflects the competition between the elastic and the active stresses: the velocity of defects arises from the reorientation of the director field due to the elastic stresses, and the advection of the director field is due to the active back flow.

Apart from the defect \emph{positions}, we have also obtained the dynamics of their \emph{orientation} in Eq. (\ref{SDO}). The definition of the defect orientation and its passive dynamics have been discussed thoroughly and adequately in Refs.~\cite{Vromans2016,Tang2017}. These authors discovered that the director field can have an extra distortion because of the relative rotation of defects. Accordingly, our form of the director field in Eq. (\ref{cubs}) is chosen such that the relative orientation of the defects is set to their optimal configuration, which means it does not have any elastic interaction that arises from extra distortions of the director field due to a relative rotation of defects. In fact, according to Refs.~\cite{Vromans2016,Tang2017}, Eq. (\ref{cubs}) implies a special choice of the boundary condition of the director field near defects, so that the relative orientation of defects is locked to a certain angle. The same is implied in Eq. (\ref{SDO}), which does not contain the Frank elastic torque. We believe this is a reasonable result in the light of our assumption that the characteristic timescale of director field relaxation is faster than that of defect motion. The activity enters the system via the back flow, according to Eq. (\ref{acfl}), and the direction of the flow is always \emph{along} the symmetry axis of the defect; hence, the flow does not impact the defect \emph{transversely}, and the \emph{elastic locking} of defects maintains the relative orientation. As a consequence, the relaxation dynamics of the relative orientation of defects is purely passive and very fast, and so we expect it to only enter at higher order in perturbation theory.

\begin{figure*}
\includegraphics[width=\linewidth]{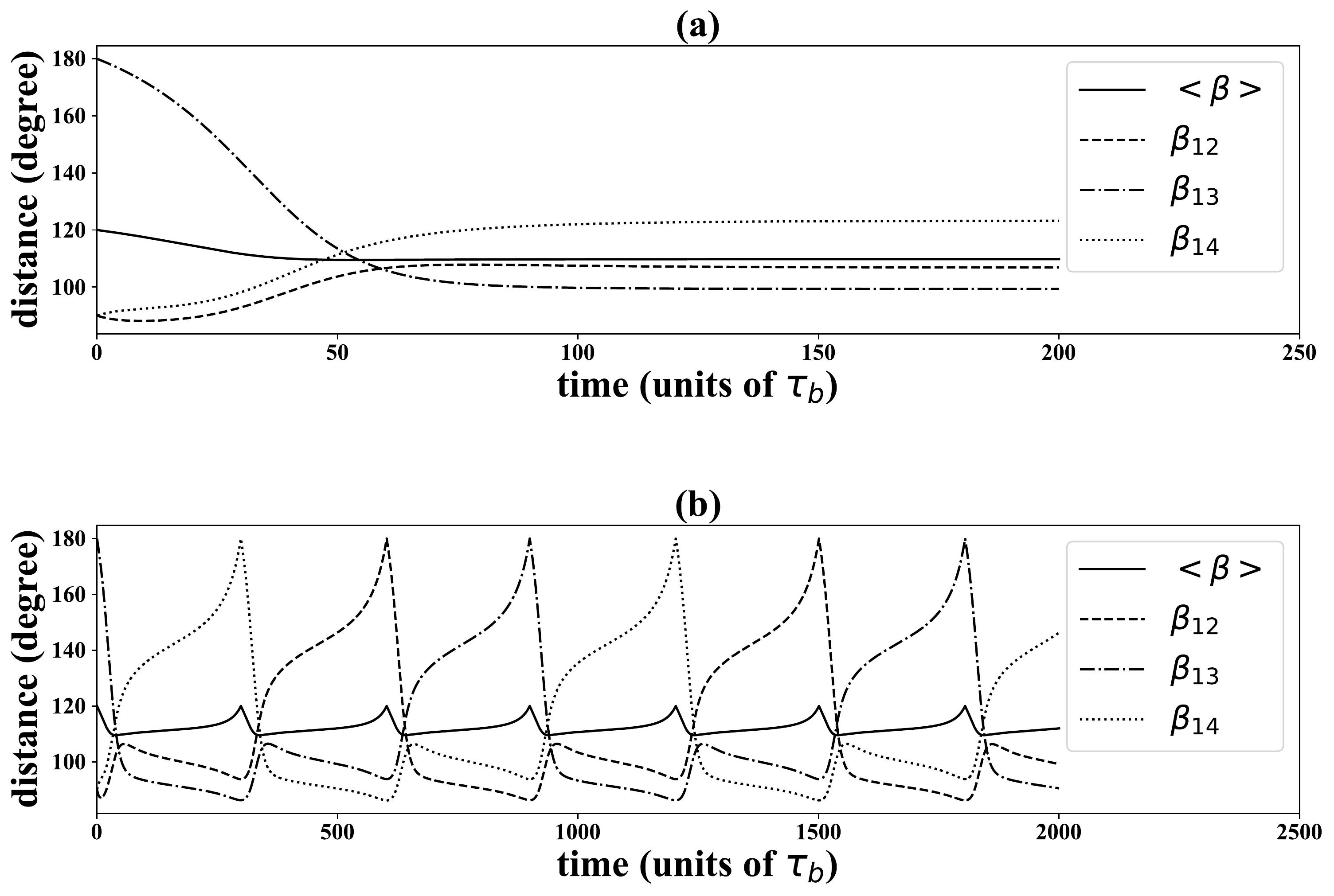}
\caption{Time evolution of the average angular distance $\langle\beta_{jk}\rangle_{jk}$ between defects at relatively low activity; recall that the planar configuration corresponds to $\langle\beta_{jk}\rangle_{jk}=\ang{120}$, while the tetrahedral configuration corresponds to $\langle\beta_{jk}\rangle_{jk}=109.5^{\circ}$. (a) For the sub-threshold activity $\alpha/\zeta=\SI{-0.16}{\micro m^2/ms}$, the defects approach stationary positions after a brief transient, reflecting the balance between the elastic interaction of defects and the active flow. (b) For the activity $\alpha/\zeta=\SI{-0.26}{\micro m^2/ms}$, the defects periodically oscillate between the tetrahedral and the planar configuration.
\label{ad}}
\end{figure*}

\begin{figure}
\includegraphics[width=\linewidth]{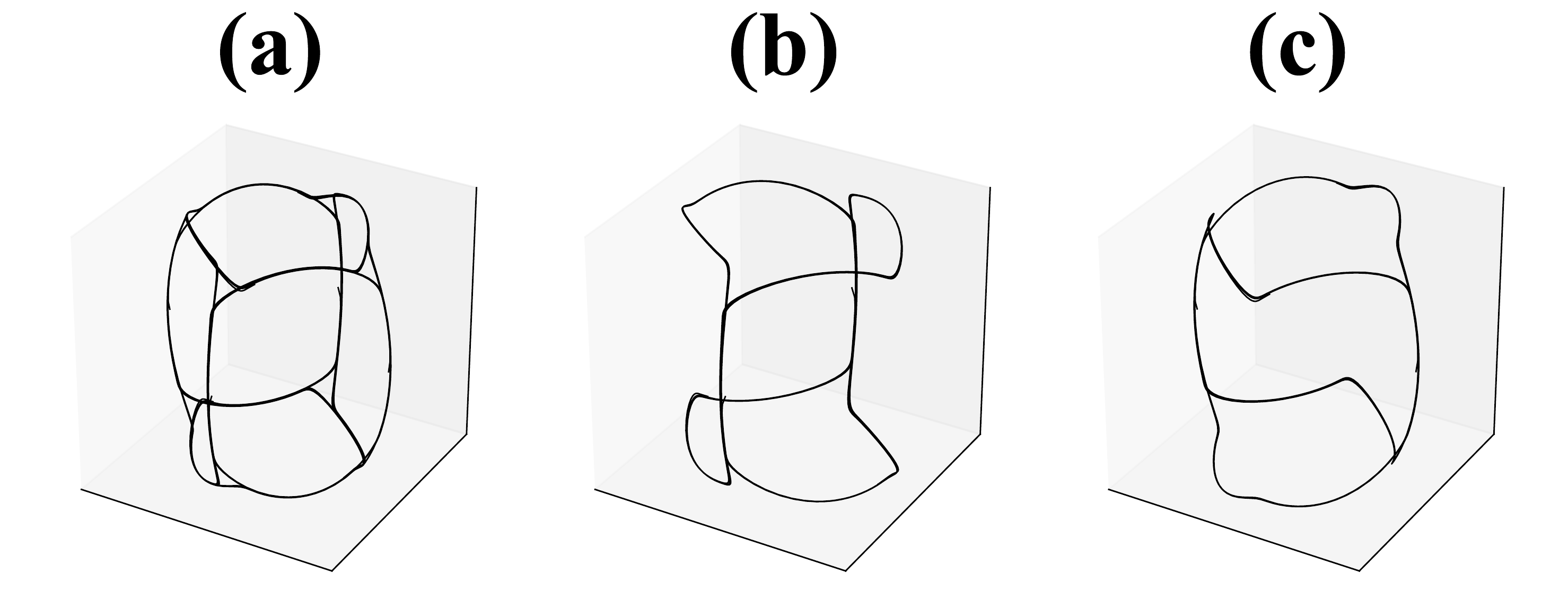}
\caption{Zigzag defect trajectories for $R/r=16$ and $\alpha/\zeta=\SI{-0.26}{\micro m^2/ms}$. The four defects trajectories in (a) decompose into two pairs of defects trajectories in (b) and (c).}
\label{tc}
\end{figure}

In order to illustrate the predictions of our theory, we will now present numerical results of the dynamic equations (\ref{SDT})--(\ref{SDO}). Here we chose $R=\SI{32}{\micro m}$, $r=\SI{2}{\micro m}$, $\gamma/2K=\SI{0.013}{ms/\micro m^2}$ and the corresponding characteristic time scale $\tau_\mathrm{b}=\gamma R^2/2K=\SI{13.312}{ms}$ as our reference baseline. The choice of the initial condition of the system we discuss is
\begin{equation}
\theta_i = \frac{\pi}{2}
\;\;,\;\;
\phi_i = \frac{\pi}{2}\times (i-1)
\;\;,\;\;
\Omega = -0.26 \ ,
\end{equation}
for $i\in\{1,2,3,4\}$. This describes four defects evenly distributed in the equatorial plane, with their direction deviating by $\ang{330.2} (i=1,3)$ or $\ang{150.2} (i=2,4)$ from the north-south direction.
We used the Python module \texttt{scipy.linalg} to numerically solve the dynamic equations (\ref{SDT}--\ref{SDO}) via the fourth-order Runge-Kutta method, using a discretization timestep of $\Delta t = 0.1\,\tau_\mathrm{b}$. Due to the huge reduction in the number of degrees of freedom (we now deal with nine first-order coupled differential equations, instead of a partial differential equation in $2+1$ dimensions), these calculations are vastly more efficient than if we had actually solved for the entire nematic field. For instance it took us only about \SI{10}{min} on a laptop with Intel Core i7-6820HQ CPU at \SI{2.70}{GHz} to complete the $2\,000\,\tau_\mathrm{b}=20\,000\,\Delta t$ trajectory shown in Fig.~\ref{ad}(b).

Our numerical results show that if the activity is below a certain threshold, the active forces are unable to continually overcome the elastic forces. After a brief dynamic transient, during which the defects re-position into an approximately tetrahedral pattern ($\langle\beta\rangle\gtrsim\ang{109.5}$), the system approaches a steady state, in which despite ongoing flow the location of defects remains stationary---see Fig.~\ref{ad}(a).

Above this threshold, our theory gives an excellent description of the periodic but highly nonlinear ratchet-like time evolution of the average angular distance $\langle\beta_{jk}\rangle_{jk}$ between defects, as shown in Fig.~\ref{ad}(b), whose characteristic pattern has been previously discovered in both numerical and experimental work \cite{Keber2014,Rui2016,Henkes2018}. The oscillation is between \ang{109.5} and \ang{120}, corresponding to a transition between the tetrahedral and the planar configuration. We also find that the dynamics of the defect orientation has a significant influence on their trajectory. Recall that in the early phenomenological theory by Khoromskaia and Alexander \cite{Khoromskaia2017} the global orientation of defects, $\Omega$, was fixed; but the authors understood the limitation of this assumption and cautioned that $\Omega$ could be important in code\-ter\-mi\-ning the motion of defects. As a result of this limitation, their theory predicted that defects approach each other in pairs. However, earlier simulation work by Zhang \emph{et al.} \cite{Rui2016} had already shown that defects do not move in pairs around their center of mass (in the regime of low activity), but follow a complex trajectory similar to what had been reported in even earlier experiments by Keber \emph{et al.} \cite{Keber2014}. We now find the same ``zig-zag'' type of complex motion in our theory, as can be seen in Fig.~\ref{tc}. An illustration for the time evolution of such a trajectory can also be seen in the Supplemental Movie \cite{sm}. We believe that the dynamics of the defect orientation $\Omega$, described in Eq. (\ref{SDO}), is responsible for this qualitative change. Indeed, if we artificially \emph{fix} the value of $\Omega$, instead of having it evolve \emph{via} Eq.~(\ref{SDO}), we observe the qualitatively different defect trajectories described by Khoromskaia and Alexander in Ref.~\cite{Khoromskaia2017}.

\begin{figure}
\includegraphics[width=\linewidth]{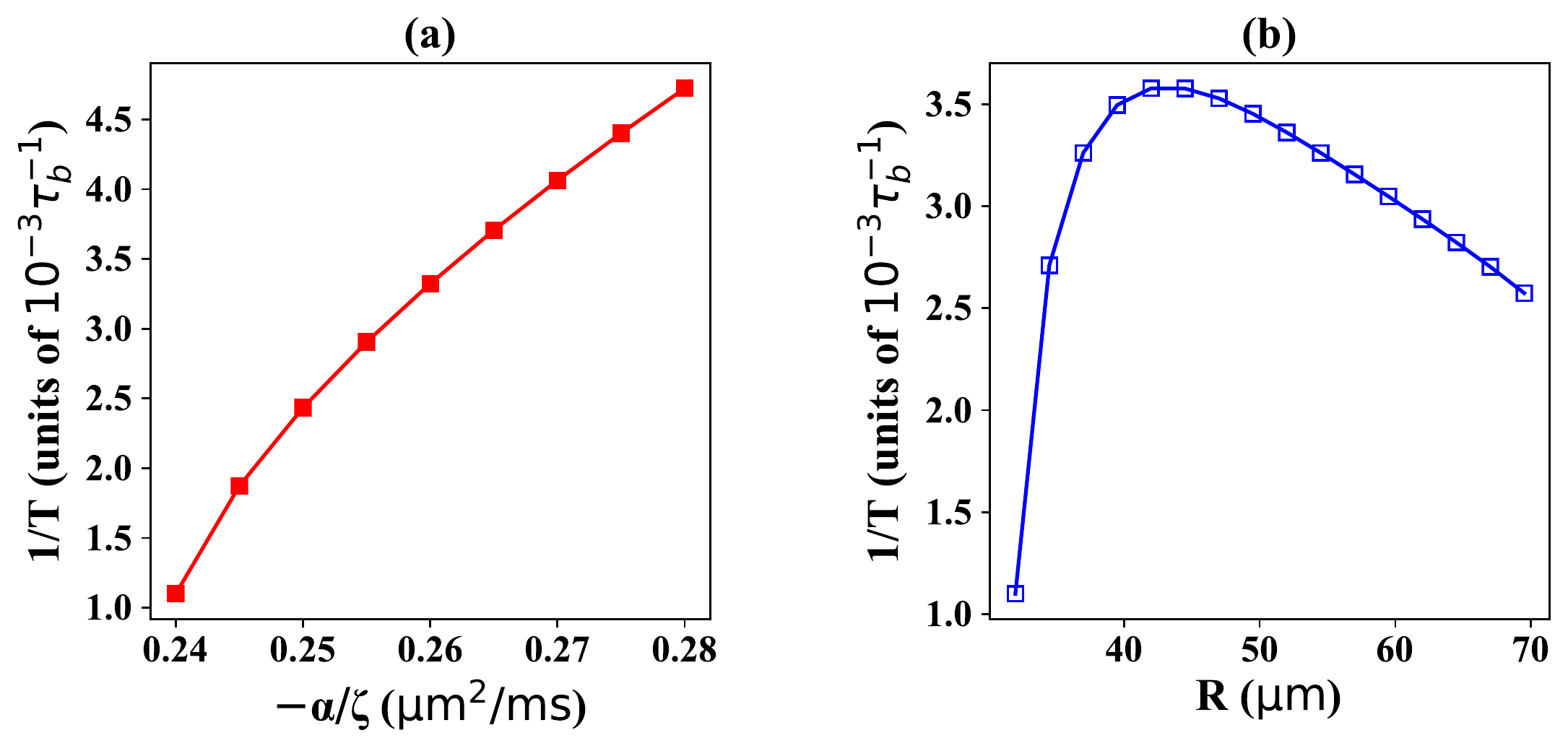}
\caption{(a) Oscillation frequency of periodic defect trajectories [such as those in Figs.~\ref{trajectory}(a) and \ref{trajectory}(d)] as a function of scaled activity $-\alpha/\zeta$ for given sphere size $R=\SI{32}{\micro m}$. (b) Same oscillation frequency, but now as a function of the substrate sphere radius $R$ for a given scaled activity of $\alpha/\zeta=\SI{-0.24}{\micro m^2/ms}$.
\label{freq}}
\end{figure}

In both experiment \cite{Keber2014} and the numerical work \cite{Rui2016} it was found that the activity tunes the frequency of the defects' oscillations between the tetrahedral and the planar configuration. We have reproduced this phenomenon in our theory. The numerical result of the dynamic equations (\ref{SDT})--(\ref{SDO}) reveals that the frequency linearly increases with increasing activity, as shown in Fig.~\ref{freq}(a). More remarkably, the size of the spherical substrate also affects the dynamics of defects. If we fix the activity of the system at $\alpha/\zeta=\SI{-0.24}{\micro m^2/ms}$, but successively increase the sphere radius $R$, the frequency of the defects' oscillatory trajectory initially increases, until it attains a maximum, beyond which it again declines, as shown in Fig.~\ref{freq}(b). To understand this, we notice that the coefficients of the mean active back flow components, (\ref{cT_j}) and (\ref{cS_j}), are both proportional to $R/r$, so increasing $R$ is equivalent to \emph{linearly} enhancing a defects' velocity. Meanwhile, the definition of characteristic timescale $\tau=\gamma R^2/2K$ suggests that a bigger $R$ will \emph{quadratically} slow the frequency of the defects' oscillation in real time, because defects have larger and larger distances to traverse on their periodic trajectories as $R$ increases. Eventually, the competition of these two effects leads to a special substrate radius at which the frequency of the defects' oscillation attains its maximum, as shown in Fig.~\ref{freq}(b).

\begin{figure*}
\includegraphics[width=\linewidth]{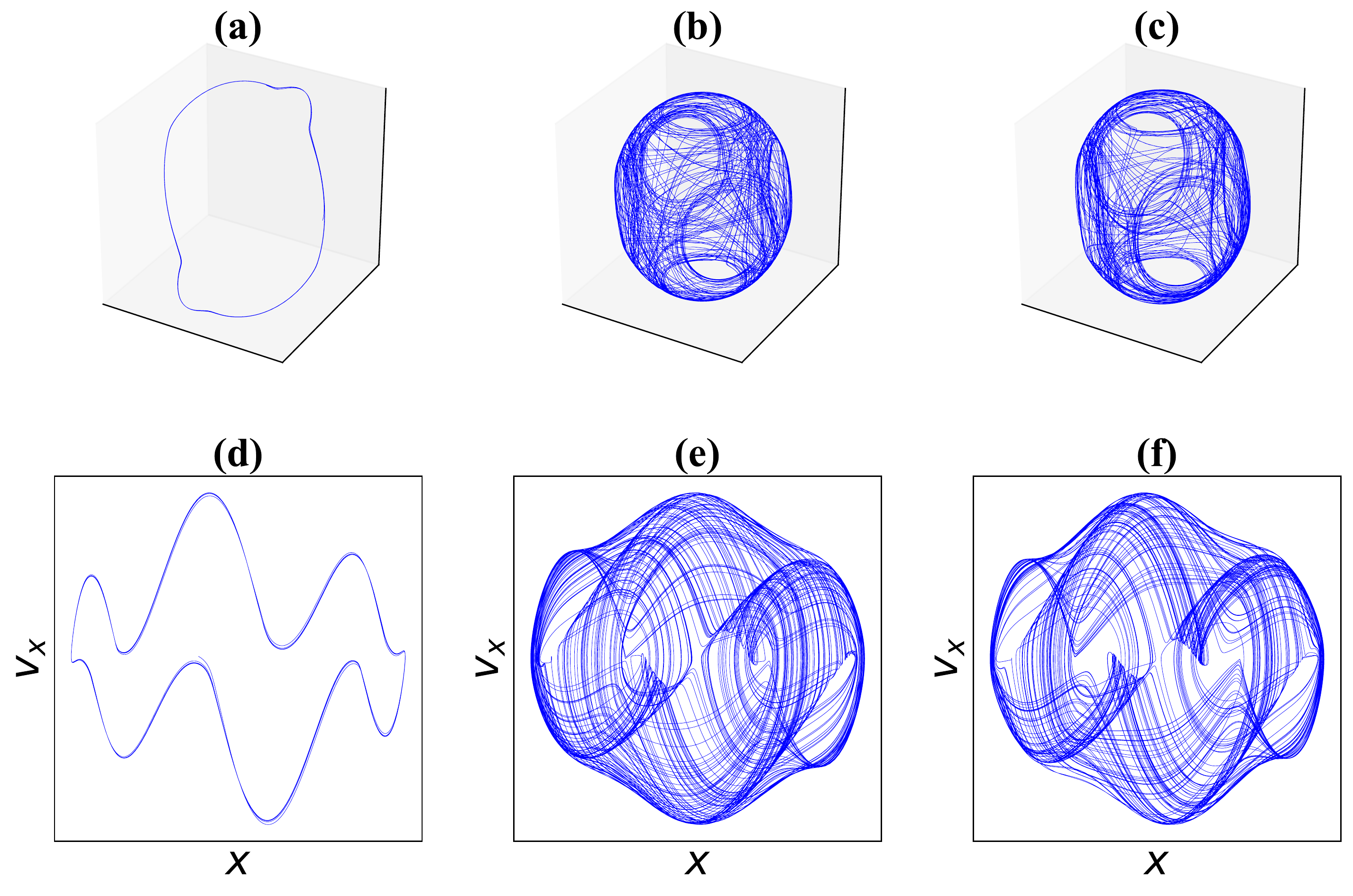}
\caption{(a--c) Trajectory of an individual defect in three dimensions. (d--f) Projection of the phase space trajectory of the same individual defect into the $X$-$V_X$ plane. Specifically: (a, d) Defect trajectories for small sphere radius $R/r=16$ and low activity $\gamma\alpha/2K\zeta=-0.00338$. (b, e) Defect trajectories for small sphere radius $R/r=16$ and high activity $\gamma\alpha/2K\zeta=-0.0208$. (c, f) Defect trajectories for large sphere radius $R/r=75$ and low activity $\gamma\alpha/2K\zeta=-0.00338$.
\label{trajectory}}
\end{figure*}

The coupling of the defect orientation and the active motion causes a rather diverse set of patterns of motion. We observed the trajectory of defects becoming chaotic when the activity or the radius of the sphere substrate is in a certain range. For low activity, the trajectory remains closed and the dynamics displays a periodic temporal evolution, as shown in Figs.~\ref{trajectory}(a) and \ref{trajectory}(d). But when we increase the activity even further, our simulations show that the dynamics will gradually enter a new regime in which the trajectories no longer close; in fact, the numerical evidence strongly suggests that the system becomes ergodic, based on the trajectory of defects in both real space and phase space [see Figs.~\ref{trajectory}(b) and \ref{trajectory}(e)], supporting the notion that at sufficiently high activity the dynamics of defects will be chaotic. And just like the chaotic transition when increasing the activity, the defect trajectory can be open for large sphere radius, as shown in Figs.~\ref{trajectory}(c) and \ref{trajectory}(f).

The chaotic transition we found in our theory is consistent with what Zhang \emph{et al.} \cite{Rui2016} observed in their significantly more challenging (Lattice-Boltzmann based) simulations for the evolution of the entire nematic field---which is potentially more prone to numerical instabilities. It is hence reassuring to see that the same transition to chaos is observed in our much simpler system of coupled ordinary differential equations.

\section{CONCLUSIONS AND OUTLOOK}

Using Onsager's variational approach to nonequilibrium (thermo-)dynamics, we have derived an effective theory for the defect dynamics in an active nematic confined to the surface of a sphere. Within our formalism the state of the system is fully determined by the positions and orientations of a small number of defects, whose dynamics follows from solving a set of coupled first-order differential equations. This is both conceptually more direct and numerically more efficient than the alternative of working with a partial differential equation for the nematic field, which lives in $(2+1)$ dimensions and describes either the full order parameter tensor or at least the nematic director.

We focused on the fundamental question how to arrive at such an effective description, and hence we have made a few approximations along the way in order to avoid cluttering the path with incidental technical difficulties. Our theory is therefore no equivalent rewriting of the original problem but rather a first-order approximation for the exact time evolution in terms of convenient smallness parameters. Nevertheless, it offers several quantitative and qualitative insights, and it efficiently reproduces previous numerical and experimental results. For instance, in the weakly driven regime we confirmed the existence of a threshold activity below which defect motion is arrested, and we characterized the shape of the defect trajectories past this threshold. In the high activity regime we observed transition into chaos. And as an example of a conceptual insight, we have elucidated the importance of promoting the global rotation $\Omega$ to a dynamic variable in order to reproduce the type of trajectories observed in experiment.


Our theoretical development can be generalized in a number of ways. Most straightforwardly, one could relax the one-con\-stant-ap\-prox\-i\-ma\-tion and investigate how differing moduli for splay and bend affect configuration and dynamics of an active nematic field with nontrivial topology \cite{Shin2008}. More challenging, in order to understand the sometimes surprisingly large fluctuations of the nematic density observed in both experiments \cite{Keber2014} and numerical studies \cite{Henkes2018}, one must include density contributions to the Rayleighian (\ref{grel}) and add the expression that follows from the associated variation to the set of dynamic equations. 

More interesting, but also much more challenging, would be steps to complete our formalism into a true fluctuating effective field theory, in which particle number is not conserved; in other words, to incorporate pair creation and annihilation events. These occur at sufficiently large activity or sphere radius and have been numerically observed \cite{Henkes2018}. The difficulty is that annihilating a $+1/2$ against a $-1/2$ defect, or creating such a pair out of an initially smooth background field, ultimately rests on the local physics our effective theory has integrated out. It might seem that at least the \emph{annihilation} event is relatively easy to account for, as one could simply eliminate a particle-antiparticle pair as soon as they approach closer than some critical distance; but formulating conditions on the local smooth field that trigger pair \emph{creation} is nontrivial. Moreover, since the rates of creation and annihilation will surely satisfy joint thermodynamic constraints, these two processes are not independent and hence cannot be treated in isolation.

Even more challenging questions arise once we look at actual experimental realizations of active nematics, such as the lipid vesicles covered by microtubule filaments that are rendered active by the addition of kinesin motors \cite{Keber2014}. The coupling between the fluid-elastic curvature energetics of the vesicle and the elasticity underlying the nematic liquid crystal impacts the \emph{morphology} of the vesicle, which need not remain a perfect sphere. In fact, protrusions growing from the location of the defects have been observed \cite{Keber2014}, and the vesicle may deform into a spin\-dle-like structure with two $+1$ defects at the spindle poles \cite{Keber2014}. Metselaar \emph{et al.} \cite{Metselaar2019} have recently proposed a continuum model for such a deformable active nematic shell. Our own framework could be extended to include this as well, by incorporating Helfrich's curvature elastic energy \cite{helfrich1973elastic} for the vesicle shape and the dissipation associated with changes of this shape into the Rayleighian (\ref{grel}). However, this will require some significant new formalism, for instance replacing the simple Riemann mapping of a sphere to a more general surface parametrization. But since throughout our development the description of the geometry is fully covariant, we suspect that the framework of Onsager's variational principle will continue to provide a convenient starting point for developing an effective theory.

\begin{acknowledgments}
We are grateful for valuable and inspiring discussions with Masao Doi and Tanniemola Liverpool. We also acknowledge financial support from the National Natural Science Foundation of China (Grants No. 11675017 and No. 11975050) and partial support from the National Science Foundation of the United States (Grant No. CHE 1764257).
\end{acknowledgments}

\appendix

\section{The Rayleighian}
According to Onsager's variational principle, only terms that explicitly contain $\partial_tQ^{ab}$ will contribute to the dynamic equations of defects. So the Rayleighian can be written as
\begin{align}
    \mathfrak{R}&=\frac{\mathrm{d}}{\mathrm{d}t}F+\frac{\gamma}{2}\int\mathrm{d}S\left(\partial_tQ^{ab}\partial_tQ_{ab}\right.\nonumber\\
    &\phantom{=}\;\left.+2\partial_tQ^{ab}V^c\nabla_cQ_{ab}-2\partial_tQ^{ab}S_{ab}\right).
\end{align}
If we substitute the active flow $V^a=\frac{\alpha}{\zeta}\nabla_bQ^{ba}$, we obtain
\begin{align}
    &\partial_tQ^{ab}S_{ab}\nonumber\\
    =&\xi q\partial_tQ^{ab}\nabla_{[a}V_{b]}+\partial_tQ^{ab}Q_a{}^c\nabla_{(c}V_{b)}-\partial_tQ^{ab}\nabla_{(a}V_{c)}Q^c{}_b\nonumber\\
    =&\frac{\alpha\xi q}{\zeta}\partial_tQ^{ab}\nabla_a\nabla_dQ^d{}_b+\frac{\alpha}{\zeta}\partial_tQ^{ab}Q_a{}^c\nabla_{(c}\nabla_dQ^d{}_{b)}\nonumber\\
    &\phantom{=}\;-\frac{\alpha}{\zeta}\partial_tQ^{ab}\nabla_{(a}\nabla_dQ^d{}_{c)}Q^c{}_b\nonumber\\
    =&\frac{\alpha}{\zeta}\partial_tQ^{ab}Q_a{}^c\nabla_{(c}\nabla_dQ^d{}_{b)}-\frac{\alpha}{\zeta}\partial_tQ^{ab}\nabla_{(a}\nabla_dQ^d{}_{c)}Q^c{}_b\nonumber\\
    &\phantom{=}\;+\frac{\alpha\xi q^3}{\zeta}\Delta\psi\partial_t\psi\label{coupling}
\end{align}
The equilibrium configuration of the director field ensures that the last term in Eqn.~(\ref{coupling}) vanishes, so the flow aligning parameter $\xi$ does not contribute. Finally, all the coupling terms read
\begin{align}
    &\frac{\gamma}{2}\int\mathrm{d}S\left(\partial_tQ^{ab}\partial_tQ_{ab}+2\partial_tQ^{ab}V^c\nabla_cQ_{ab}-2\partial_tQ^{ab}S_{ab}\right)\nonumber\\
    =&\gamma q^2\int\mathrm{d}S\left(\partial_t\psi\right)^2-\frac{2\gamma\alpha q^2}{\zeta}\int\mathrm{d}S\nabla_b\left(\nabla_a\psi-A_a\right)Q^{ab}\partial_t\psi.\label{Ray}
\end{align}

\section{Riemann sphere representation}
Before calculating integrals (\ref{M_jk})-(\ref{S_j}), we must express each term in the Riemann sphere representation by using the stereographic projection $z(\theta, \phi)=2R\tan{\frac{\theta}{2}}\mathrm{e}^{\mathrm{i}\phi}$ shown in the main text. This leads to
\begin{align}
    \int\mathrm{d}S&=\int\mathrm{d}z\mathrm{d}{\bar{z}}\frac{1}{2\left(1+\frac{z\bar{z}}{4R^2}\right)^2}\label{rint}\\
    \partial_{\theta_k}\psi_k&=\frac{\mathrm{i}}{4}R\left(1+\frac{z_k\bar{z}_k}{4R^2}\right)\sqrt{\frac{z_k}{\bar{z}_k}}\frac{1}{z-z_k}+\cc,\label{rthe}\\
    \partial_{\phi_k}\psi_k&=-\frac14\frac{z_k}{z-z_k}+\cc,\label{rphi}\\
    \partial^2_\phi\psi_k&=\frac{\mathrm{i}}{4}\frac{z}{z-z_k}-\frac{\mathrm{i}}{4}\frac{z^2}{(z-z_k)^2}+\cc,\label{rphi2}\\
    \partial_{\theta}\partial_{\phi}\psi_k&=\frac R4\left(1+\frac{z\bar{z}}{4R^2}\right)\sqrt{\frac{z}{\bar{z}}}\frac{1}{z-z_k}\nonumber\\
    &\phantom{=}\;-\frac R4\left(1+\frac{z\bar{z}}{4R^2}\right)\sqrt{\frac{z}{\bar{z}}}\frac{z}{(z-z_k)^2}+\cc,\label{rthp}\\
    \partial^2_\theta\psi_k&=-\frac{\mathrm{i}}{8}R\left(1+\frac{z\bar{z}}{4R^2}\right)^2\frac{4R\sqrt{z\bar z}}{4R^2+z\bar z}\sqrt{\frac{z}{\bar{z}}}\frac{1}{z-z_k}\nonumber\\
    &\phantom{=}\;+\frac{\mathrm{i}}{4}R^2\left(1+\frac{z\bar{z}}{4R^2}\right)^2\frac{z}{\bar{z}}\frac{1}{(z-z_k)^2}+\cc \ . \label{rthe2}
\end{align}

We can calculate the area integrals (\ref{M_jk}-\ref{S_j}) based on the following theorem:

For a given function $G$ in an area $S$ with the boundary $\partial S$ in the complex plane $z=x+\mathrm{i}y$,
\begin{align}
    \int_S\partial_zG(z,\bar{z})\,\mathrm{d}z\mathrm{d}\bar{z}
    &=\int_S2\partial_zG(z,\bar{z})\,\mathrm{d}x\mathrm{d}y\nonumber\\
    &=\int_S(\partial_x-\mathrm{i}\partial_y)G\,\mathrm{d}x\mathrm{d}y \ .\label{green1}
\end{align}
Based on Green's theorem, if the first derivative of $G$ is continuous in $S$, then we have
\begin{align}
    \int_S\partial_zG(z,\bar{z})\,\mathrm{d}z\mathrm{d}\bar{z}
    &=\int_S(\partial_x-\mathrm{i}\partial_y)G\,\mathrm{d}x\mathrm{d}y\nonumber\\
    &=\oint_{\partial S}(G\mathrm{d}y+\mathrm{i}G\,\mathrm{d}x)\nonumber\\
    &=\mathrm{i}\oint_{\partial S} G\,\mathrm{d}\bar{z}.\label{green}
\end{align}
Then we can use
\begin{align}
    \int_S\partial_zG(z,\bar{z})\,\mathrm{d}z\mathrm{d}\bar{z}=\mathrm{i}\oint_{\partial S} G\,\mathrm{d}\bar{z}\label{cgreen}
\end{align}
to transform an area integral into a boundary integral. And if $G$ is not analytic, the integral will depend on the shape of the contour of the boundary, instead of just the topology.

\section{Mobility matrices}
In order to obtain the result of integral in (\ref{M_jk}-\ref{Phi_j}), there are two basic integrals we need to calculate first:
\begin{align}
    &\int\mathrm{d}S\frac{1}{\left(z-z_j\right)\left(z-z_k\right)} \ ,\label{core1}\\
    &\int\mathrm{d}S\frac{1}{\left(\bar{z}-\bar{z}_j\right)\left(z-z_k\right)}\label{core2} \ .
\end{align}
According to Eq. (\ref{rint})-(\ref{rphi}), the mobility matrices (\ref{M_jk})-(\ref{Phi_j}) can be considered as linear combinations of the integrals (\ref{core1}) and (\ref{core2}) with their complex conjugates. The boundaries $\partial S$ are those peripheries around each defects based on the integral area we require. First, we use Eq. (\ref{cgreen}) to transform the area integrals into line integrals at each boundaries. Then we expand the integrand near each boundary, and only keep the first order based on the assumption $R\gg r$. The detailed procedure is demonstrated below.

\subsection{The calculation of the integral (\ref{core1})}
\subsubsection{Diagonal element: \texorpdfstring{$k=j$}{Lg}}
\begin{align}
    &\int\mathrm{d}z\mathrm{d}\bar{z}\frac{1}{2 \left(1+\frac{z\bar{z}}{4R^2}\right)^2\left(z-z_j\right)^2}\nonumber\\
    =&\mathrm{i}\oint_C\mathrm{d}z\frac{8R^4}{z \left(4R^2+z\bar{z}\right)\left(z-z_j\right)^2}\nonumber\\
    =&\mathrm{i}\oint_{C_0+C_j}\mathrm{d}z\frac{-8R^4}{z \left(4R^2+z\bar{z}\right)\left(z-z_j\right)^2}\nonumber\\
    =&\mathrm{i}\oint_{C_0}\mathrm{d}z\frac{-2R^2}{z \left(z-z_j\right)^2}+\mathrm{i}\oint_{C_j}\mathrm{d}z\frac{-8R^4}{z \left(4R^2+z\bar{z}_j\right)\left(z-z_j\right)^2}\nonumber\\
    =&\frac{4\pi R^2\bar{z}_j^2}{\left(4R^2+z_j\bar{z}_j\right)^2}\label{zzj2}
\end{align}
In order to fulfill the condition required by Eq. (\ref{cgreen}), we need to exclude the origin of the complex plane so that the integrand is continuous over the domain of integration. Then the boundary $C$ should contain the boundary $C_0$ around the origin, and the boundary $C_j$ around the defect at $z_j$. And with performing the same procedure, the nondiagonal element can be calculated as follows.

\subsubsection{Nondiagonal element: \texorpdfstring{$k\neq j$}{Lg}}
\begin{align}
    &\int\mathrm{d}z\mathrm{d}\bar{z}\frac{1}{2 \left(1+\frac{z\bar{z}}{4R^2}\right)^2\left(z-z_j\right)\left(z-z_k\right)}\nonumber\\
    =&\mathrm{i}\oint_C\mathrm{d}z\frac{8R^4}{z \left(4R^2+z\bar{z}\right)\left(z-z_j\right)\left(z-z_k\right)}\nonumber\\
    =&\mathrm{i}\oint_{C_0+C_j+C_k}\mathrm{d}z\frac{-8R^4}{z \left(4R^2+z\bar{z}\right)\left(z-z_j\right)\left(z-z_k\right)}\nonumber\\
    =&\mathrm{i}\oint_{C_0}\mathrm{d}z\frac{-2R^2}{z \left(z-z_j\right)\left(z-z_k\right)}\nonumber\\
    &\phantom{=}\;+\mathrm{i}\oint_{C_j}\mathrm{d}z\frac{-8R^4}{z \left(4R^2+z\bar{z}_j\right)\left(z-z_j\right)\left(z-z_k\right)}\nonumber\\
    &\phantom{=}\;+\mathrm{i}\oint_{C_k}\mathrm{d}z\frac{-8R^4}{z \left(4R^2+z\bar{z}_k\right)\left(z-z_j\right)\left(z-z_k\right)}\nonumber\\
    =&\frac{16\pi R^4(\bar{z}_k-\bar{z}_j)-4\pi R^2\bar{z}_j\bar{z}_k \left(z_k-z_j\right)}{\left(z_j-z_k\right)\left(4R^2+z_j\bar{z}_j\right)\left(4R^2+z_k\bar{z}_k\right)}\label{zzjk}
\end{align}

\subsection{The calculation of the integral (\ref{core2})}
\subsubsection{Diagonal element: \texorpdfstring{$k=j$}{Lg}}
\begin{align}
    &\int\mathrm{d}z\mathrm{d}\bar{z}\frac{1}{2 \left(1+\frac{z\bar{z}}{4R^2}\right)^2\left(\bar{z}-\bar{z}_j\right)\left(z-z_j\right)}\nonumber\\
    =&-\mathrm{i}\oint_{C}\mathrm{d}z \left[\frac{8R^4}{\left(4R^2+z\bar{z}\right)\left(4R^2+z\bar{z}_j\right)\left(z-z_j\right)}\right.\nonumber\\
    &\left.+\frac{8R^4}{\left(4R^2+z\bar{z}_j\right)^2\left(z-z_j\right)}\ln\frac{z\bar{z}-z\bar{z}_j}{4R^2+z\bar{z}}\right]\nonumber\\
    =&-\mathrm{i}\oint_{C}\mathrm{d}z \left[\frac{8R^4}{\left(4R^2+z\bar{z}\right)\left(4R^2+z\bar{z}_j\right)\left(z-z_j\right)}\right.\nonumber\\
    &\left.+\frac{8R^4}{\left(4R^2+z\bar{z}_j\right)^2\left(z-z_j\right)}\ln\frac{\left(z-z_j\right)\left(\bar{z}-\bar{z}_j\right)}{4R^2+z\bar{z}}\right]\nonumber\\
    &-\mathrm{i}\oint_{C}\mathrm{d}z\frac{8R^4}{\left(4R^2+z\bar{z}_j\right)^2\left(z-z_j\right)}\ln\frac{z}{z-z_j}\nonumber\\
    =&-\mathrm{i}\oint_{C}\mathrm{d}z \left[\frac{8R^4}{\left(4R^2+z\bar{z}\right)\left(4R^2+z\bar{z}_j\right)\left(z-z_j\right)}\right.\nonumber\\
    &\left.+\frac{8R^4}{\left(4R^2+z\bar{z}_j\right)^2\left(z-z_j\right)}\ln\frac{\left(z-z_j\right)\left(\bar{z}-\bar{z}_j\right)}{4R^2+z\bar{z}}\right]\nonumber\\
    =&\mathrm{i}\oint_{C_0}\mathrm{d}z\left[\frac{2R^2z\bar{z}_j}{\left(4R^2+z\bar{z}_j\right)^2 \left(z-z_j\right)}\right.\nonumber\\
    &\left.+\frac{8R^4}{\left(4R^2+z\bar{z}_j\right)^2 \left(z-z_j\right)}\left(\ln\frac{z_j\bar{z}_j-z\bar{z}_j}{4R^2}+1\right)\right]\nonumber\\
    &+\mathrm{i}\oint_{C_j}\mathrm{d}z\frac{8R^4}{\left(4R^2+z\bar{z}_j\right)^2\left(z-z_j\right)}\left(\ln\frac{r_j^2}{4R^2+z\bar{z}_j}+1\right)\nonumber\\
    &+\mathrm{i}\oint_{C_j^*}\mathrm{d}z\left[\frac{-2R^2z_j}{\left(4R^2+z\bar{z}_j\right)\left(z-z_j\right)^2}\right.\nonumber\\
    &\left.+\frac{8R^4}{\left(4R^2+z\bar{z}_j\right)^2\left(z-z_j\right)}\ln\frac{4R^2+z_j\bar{z}_j}{4R^2}\right]\nonumber\\
    =&\frac{4\pi R^2z_j\bar{z}_j}{\left(4R^2+z_j\bar{z}_j\right)^2}-\frac{16\pi R^4}{\left(4R^2+z_j\bar{z}_j\right)^2}\left(1+2\ln\frac{2Rr_j}{4R^2+z_j\bar{z}_j}\right)\label{czj2}
\end{align}

\begin{figure}
\includegraphics[scale=0.7]{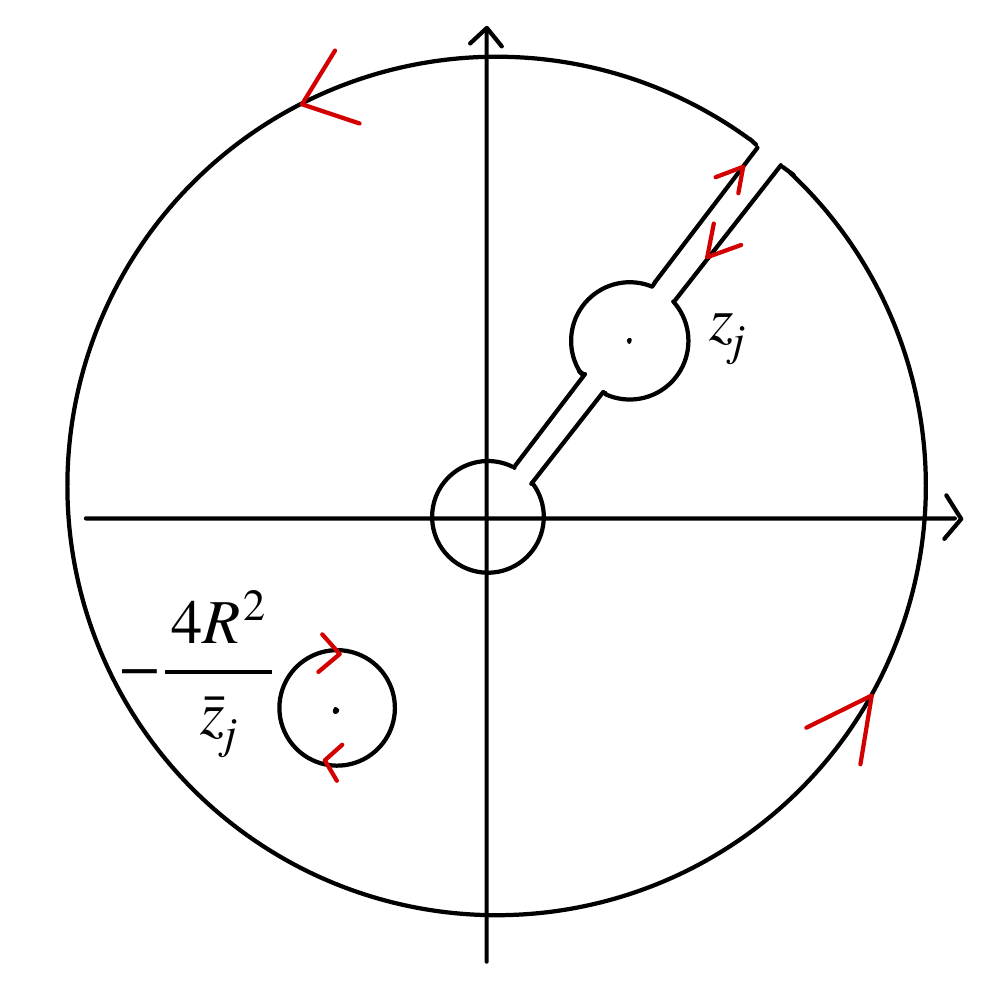}
\caption{Integral contour of (\ref{czj2}).\label{contour}}
\end{figure}

The integrand indicates that there are three singular points: $z=0$, $z=z_j$, and $z=-4R^2/\bar{z}_j$, corresponding to boundaries of $C_0$, $C_j$ and $C^*_j$ separately. In addition, the first two are also branch points because of the logarithmic function. In order to use Eq. (\ref{cgreen}), we again need the integrand to be continuous along the boundary. We therefore add a branch cut at both of the boundary $C_0$ around the branch point $z=0$, and the boundary $C_j$ around the branch point $z=z_j$. Once we take the boundary $C=C_0+C_j+C^*_j$ as shown in Fig.(\ref{contour}), the line integral of the sixth line is zero, according to the residue theorem.

\subsubsection{Non-diagonal element: \texorpdfstring{$k\neq j$}{Lg}}
\begin{align}
    &\int\mathrm{d}z\mathrm{d}\bar{z}\frac{1}{2 \left(1+\frac{z\bar{z}}{4R^2}\right)^2\left(\bar{z}-\bar{z}_j\right)\left(z-z_k\right)}\nonumber\\
    =&-\mathrm{i}\oint_{C}\mathrm{d}z \left[\frac{8R^4}{\left(4R^2+z\bar{z}\right)\left(4R^2+z\bar{z}_j\right)\left(z-z_k\right)}\right.\nonumber\\
    &\left.+\frac{8R^4}{\left(4R^2+z\bar{z}_j\right)^2\left(z-z_k\right)}\ln\frac{z\bar{z}-z\bar{z}_j}{4R^2+z\bar{z}}\right]\nonumber\\
    =&\mathrm{i}\oint_{C_0}\mathrm{d}z\left[\frac{2R^2z\bar{z}_j}{\left(4R^2+z\bar{z}_j\right)^2 \left(z-z_k\right)}\right.\nonumber\\
    &\left.+\frac{8R^4}{\left(4R^2+z\bar{z}_j\right)^2 \left(z-z_k\right)}\left(\ln\frac{z_j\bar{z}_j-z\bar{z}_j}{4R^2}+1\right)\right]\nonumber\\
    &+\mathrm{i}\oint_{C_j}\mathrm{d}z\frac{8R^4}{\left(4R^2+z\bar{z}_j\right)^2 \left(z-z_k\right)}\left(\ln\frac{r_j^2}{4R^2+z\bar{z}_j}+1\right)\nonumber\\
    &+\mathrm{i}\oint_{C_k}\mathrm{d}z\left[\frac{8R^4}{\left(4R^2+z\bar{z}_j\right)\left(4R^2+z\bar{z}_k\right)\left(z-z_k\right)}\right.\nonumber\\
    &\left.+\frac{8R^4}{\left(4R^2+z\bar{z}_j\right)^2\left(z-z_k\right)}\ln\frac{\left(z-z_j\right)\left(\bar{z}_k-\bar{z}_j\right)}{4R^2+z\bar{z}_k}\right]\nonumber\\
    &+\mathrm{i}\oint_{C_j^*}\mathrm{d}z\left[\frac{-2R^2z_j}{\left(4R^2+z\bar{z}_j\right)\left(z-z_j\right)\left(z-z_k\right)}\right.\nonumber\\
    &\left.+\frac{8R^4}{\left(4R^2+z\bar{z}_j\right)^2\left(z-z_k\right)}\ln\frac{4R^2+z_j\bar{z}_j}{4R^2}\right]\nonumber\\
    =&\frac{-4\pi R^2}{\left(4R^2+z_k\bar{z}_j\right)^2}\left[\frac{\left(4R^2+z_k\bar{z}_j\right)\left(16R^4-z_j\bar{z}_jz_k\bar{z}_k\right)}{\left(4R^2+z_j\bar{z}_j\right)\left(4R^2+z_k\bar{z}_k\right)}\right.\nonumber\\
    &\left.+4R^2\ln\frac{4R^2 \left(z_j-z_k\right)\left(\bar{z}_j-\bar{z}_k\right)}{\left(4R^2+z_j\bar{z}_j\right)\left(4R^2+z_k\bar{z}_k\right)}\right]\label{czjk1}
\end{align}
Notice that the result seems to be divergent when $z_k=-4R^2/\bar{z}_j$. This is because we treat $z_k$ and $-4R^2/\bar{z}_j$ as two different singular points when performing the residue theorem in the calculation. If we instead consider the case of $z_k=-4R^2/\bar{z}_j$, we have:
\begin{align}
    &\int\mathrm{d}z\mathrm{d}\bar{z}\frac{1}{2 \left(1+\frac{z\bar{z}}{4R^2}\right)^2\left(\bar{z}-\bar{z}_j\right)\left(z-z_k\right)}\nonumber\\
    =&\int\mathrm{d}z\mathrm{d}\bar{z}\frac{1}{2 \left(1+\frac{z\bar{z}}{4R^2}\right)^2\left(\bar{z}-\bar{z}_j\right)\left(z+\frac{4R^2}{\bar{z}_j}\right)}\nonumber\\
    =&-\mathrm{i}\oint_{C}\mathrm{d}z \left[\frac{8R^4\bar{z}_j}{\left(4R^2+z\bar{z}\right)\left(4R^2+z\bar{z}_j\right)^2}\right.\nonumber\\
    &\left.+\frac{8R^4\bar{z}_j}{\left(4R^2+z\bar{z}_j\right)^3}\ln\frac{z\bar{z}-z\bar{z}_j}{4R^2+z\bar{z}}\right]\nonumber\\
    =&\mathrm{i}\oint_{C_0}\mathrm{d}z\left[\frac{2R^2z\bar{z}_j^2}{\left(4R^2+z\bar{z}_j\right)^3}\right.\nonumber\\
    &\left.+\frac{8R^4\bar{z}_j}{\left(4R^2+z\bar{z}_j\right)^3}\left(\ln\frac{z_j\bar{z}_j-z\bar{z}_j}{4R^2}+1\right)\right]\nonumber\\
    &+\mathrm{i}\oint_{C_j}\mathrm{d}z\frac{8R^4\bar{z}_j}{\left(4R^2+z\bar{z}_j\right)^3}\left(\ln\frac{r_j^2}{4R^2+z\bar{z}_j}+1\right)\nonumber\\
    &+\mathrm{i}\oint_{C_j^*}\mathrm{d}z\left[\frac{-2R^2z_j\bar{z}_j}{\left(4R^2+z\bar{z}_j\right)^2\left(z-z_j\right)}\right.\nonumber\\
    &\left.+\frac{8R^4\bar{z}_j}{\left(4R^2+z\bar{z}_j\right)^3}\ln\frac{4R^2+z_j\bar{z}_j}{4R^2}\right]\nonumber\\
    =&\frac{-4\pi R^2z_j\bar{z}_j}{\left(4R^2+z_j\bar{z}_j\right)^2}\label{czjk2}.
\end{align}
We can check that the result in (\ref{czjk2}) indeed coincides with (\ref{czjk1}) in the limit in which $z_k$ approaches $-4R^2/\bar{z}_j$. In fact, the singular point $-4R^2/\bar{z}_j$ is the antipode of the defect $z_j$ on the sphere, and the singularity does not have any physical meaning.

\subsection{Results of mobility matrices}

Substituting the corresponding coefficients according to Eqs. (\ref{rthe}) and (\ref{rphi}), the elements of the mobility matrices can be constructed as follows:

\textit{Diagonal element:}
\begin{align}
    P_{jj}&=\frac{\pi R^2|z_j|^4}{(4R^2+|z_j|^2)^2}\nonumber\\
    &\phantom{=}\;-\frac{2\pi R^4|z_j|^2}{(4R^2+|z_j|^2)^2}\left(1+2\ln\frac{2Rr_j}{4R^2+|z_j|^2}\right) \ ,\\
    M_{jj}&=-\frac{\pi R^2}{8}\left(1+2\ln\frac{2Rr_j}{4R^2+|z_j|^2}\right) \ .
\end{align}

\textit{Nondiagonal element:}
\begin{align}
    P_{jk}&=-\frac{\pi R^2 z_k\bar{z}_j}{4(4R^2+z_k\bar{z}_j)^2}\left[\frac{\left(4R^2+z_k\bar{z}_j\right)\left(16R^4-|z_j|^2|z_k|^2\right)}{\left(4R^2+|z_j|^2\right)\left(4R^2+|z_k|^2\right)}\right.  \nonumber\\
    &\phantom{=}\;\left.+4R^2\ln\frac{4R^2|z_j-z_k|^2}{\left(4R^2+|z_j|^2\right)\left(4R^2+|z_k|^2\right)}\right]\nonumber\\
    &\phantom{=}\;+\frac{\pi R^2z_jz_k}{4(z_j-z_k)}\left(\frac{\bar{z}_k}{4R^2+|z_k|^2}-\frac{\bar{z}_j}{4R^2+|z_j|^2}\right)+\cc \ . \\
    M_{jk}&=\frac{4\pi R^2 \left(\bar{z}_j-\bar{z}_k\right)+\pi\bar{z}_k\bar{z}_j \left(z_k-z_j\right)}{64 \left(z_j-z_k\right)}\sqrt{\frac{z_jz_k}{\bar{z}_j\bar{z}_k}}\nonumber\\
    &\phantom{=}\;-\frac{\pi \left(16R^4-|z_j|^2|z_k|^2\right)}{64\left(4R^2+z_k\bar{z}_j\right)}\sqrt{\frac{\bar{z}_jz_k}{z_j\bar{z}_k}}\nonumber\\
    &\phantom{=}\;-\frac{\pi R^2\left(4R^2+|z_j|^2\right)\left(4R^2+|z_k|^2\right)}{16 \left(4R^2+z_k\bar{z}_j\right)^2}\sqrt{\frac{\bar{z}_jz_k}{z_j\bar{z}_k}}\nonumber\\
    &\phantom{=}\;\ln\frac{4R^2|z_j-z_k|^2}{\left(4R^2+|z_j|^2\right)\left(4R^2+|z_k|^2\right)}+\cc \ ,\\
    N_{jk}&=-\mathrm{i}\pi Rz_k\sqrt{\frac{z_j}{\bar{z}_j}}\left[\frac{R^2 \left(\bar{z}_k-\bar{z}_j\right)}{4 \left(z_j-z_k\right)\left(4R^2+|z_k|^2\right)}\right.\nonumber\\
    &\phantom{=}\;\left.+\frac{\bar{z}_j\bar{z}_k}{16 \left(4R^2+|z_k|^2\right)}\right]\nonumber\\
    &\phantom{=}\;+\frac{\mathrm{i}\pi R\bar{z}_k \left(16R^4-|z_j|^2|z_k|^2\right)}{16\left(4R^2+z_j\bar{z}_k\right)\left(4R^2+|z_k|^2\right)}\sqrt{\frac{z_j}{\bar{z}_j}}\nonumber\\
    &\phantom{=}\;+\frac{\mathrm{i}\pi R^3\bar{z}_k \left(4R^2+|z_j|^2\right)}{4\left(4R^2+z_j\bar{z}_k\right)^2}\sqrt{\frac{z_j}{\bar{z}_j}}\nonumber\\
    &\phantom{=}\;\ln\frac{4R^2|z_j-z_k|^2}{\left(4R^2+|z_j|^2\right)\left(4R^2+|z_k|^2\right)}+\cc \ .
\end{align}
It is also easy to calculate the vector $\Phi_j$ and $\Theta_j$:
\begin{align}
    \Phi_j&=\int\mathrm{d}S\partial_{\phi_j}\psi_j\nonumber\\
    &=-\int\mathrm{d}z\mathrm{d}\bar{z}\frac{2R^4z_j}{(4R^2+z\bar{z})^2(z-z_j)}+\cc\nonumber\\
    &=-\mathrm{i}\oint_C\mathrm{d}z\frac{2R^4z_j}{z(4R^2+z\bar{z})(z-z_j)}\nonumber\\
    &=\mathrm{i}\oint_{C_0}\mathrm{d}z\frac{R^2z_j}{2z^2-2zz_j}\nonumber\\
    &\phantom{=}\;+\mathrm{i}\oint_{C_j}\mathrm{d}z\frac{2R^4z_j}{z(z-z_j)(4R^2+z\bar{z}_j)}+\cc\nonumber\\
    &=\frac{2\pi R^2|z_j|^2}{4R^2+|z_j|^2} \ ,\label{cPhi_j}\\
    \Theta_j&=\int\mathrm{d}S\partial_{\theta_j}\psi_j\nonumber\\
    &=\int\mathrm{d}z\mathrm{d}\bar{z}\frac{\mathrm{i}R^3\left(4R^2+z_j\bar{z}_j\right)}{2(4R^2+z\bar{z})^2(z-z_j)}\sqrt{\frac{z_j}{\bar{z}_j}}+\cc\nonumber\\
    &=-\oint_C\mathrm{d}z\frac{R^3\left(4R^2+z_j\bar{z}_j\right)}{2z(4R^2+z\bar{z})(z-z_j)}\sqrt{\frac{z_j}{\bar{z}_j}}\nonumber\\
    &=\oint_{C_0}\mathrm{d}z\frac{R\left(4R^2+z_j\bar{z}_j\right)}{8z(z-z_j)}\sqrt{\frac{z_j}{\bar{z}_j}}\nonumber\\
    &\phantom{=}\;+\oint_{C_j}\mathrm{d}z\frac{R^3\left(4R^2+z_j\bar{z}_j\right)}{2z(z-z_j)\left(4R^2+z\bar{z}_j\right)}\sqrt{\frac{z_j}{\bar{z}_j}}+\cc\nonumber\\
    &=-\mathrm{i}\frac{\pi}{4}R|z_j|+\mathrm{i}\frac{\pi}{4}R|z_j|=0 \ .\label{cThe_j}
\end{align}

\section{The calculation of \texorpdfstring{$T_j$, $R_j$, and $L$}{Lg}}

Here we calculate $T_j$ as an example to illustrate the procedure:
\begin{align}
    T_j&=-\frac{q_0\alpha}{2R^2\zeta}\int\mathrm{d}S\left[\sin2\psi\left(2\csc^2\theta-1\right.\right.\nonumber\\
    &\phantom{=}\;\left.\left.+2\cot\theta\csc\theta\partial_\phi\psi-2\csc\theta\partial_\theta\partial_\phi\psi\right)\right.\nonumber\\
    &\phantom{=}\;\left.-\cos2\psi\left(\partial^2_\theta\psi-\csc^2\theta\partial^2_\phi\psi-\cot\theta\partial_\theta\psi\right)\right]\partial_{\theta_j}\psi_j \ .\label{exRJ}
\end{align}
Under the Riemann sphere representation, we have
\begin{align}
    \sin2\psi&=-\frac{\mathrm{i}}{2}\exp{\left(2\mathrm{i}\Omega\right)}\frac{\bar{z}}{z}\prod_{p=1}^4\sqrt{\frac{z-z_p}{\bar{z}-\bar{z}_p}}+\cc \ ,\\
    \cos2\psi&=\frac{\mathrm{1}}{2}\exp{\left(2\mathrm{i}\Omega\right)}\frac{\bar{z}}{z}\prod_{p=1}^4\sqrt{\frac{z-z_p}{\bar{z}-\bar{z}_p}}+\cc \ .
\end{align}

It seems very hard to find a global function $G$, so that $T_j=\mathrm{i}\oint\mathrm{d}\bar{z}\,G$. Our strategy is to expand Eq. (\ref{exRJ}) near each of the defects, and subsequently do the procedure we performed above. According to Eqs. (\ref{rint})-(\ref{rthe2}), we can expand the integrand of $T_j$ in terms of $z-z_j$ in the following form:
\begin{align}
    T_j&=\mathrm{i}\oint_{C_j}\mathrm{d}\bar{z}\int\mathrm{d}z\sqrt{\frac{z-z_j}{\bar{z}-\bar{z}_j}}\nonumber\\
    &\phantom{=}\;\sum_{q=1}^3\sum_{m=0}^q\left[\frac{1}{(z-z_j)^m(\bar{z}-\bar{z}_j)^{q-m}}\sum_{l=0}^\infty A_{m,q,l}(z-z_j)^l\right]\nonumber\\
    &\phantom{=}\;+\mathrm{i}\sum_{k\neq j}\oint_{C_k}\mathrm{d}\bar{z}\int\mathrm{d}z\sqrt{\frac{z-z_k}{\bar{z}-\bar{z}_k}}\nonumber\\
    &\phantom{=}\;\sum_{n=0}^2\left[\left(\frac{1}{(z-z_k)^n}+\frac{1}{(\bar{z}-\bar{z}_k)^n}\right)\sum_{l=0}^\infty B_{n,l}(z-z_k)^l\right]+\cc\nonumber\\
    &=\mathrm{i}\oint_{C_j}\mathrm{d}\bar{z}\sum_{q=1}^3\sum_{m=0}^q\sum_{l=0}^\infty \frac{2A_{m,q,l}r_j^{3-2m+2l}}{3-2m+2l}\left(\bar{z}-\bar{z}_j\right)^{2m-l-q-2}\nonumber\\
    &\phantom{=}\;+\mathrm{i}\sum_{k\neq j}\oint_{C_k}\mathrm{d}\bar{z}\sum_{n=0}^2\sum_{l=0}^\infty B_{n,l}\left[\frac{2r_k^{3+2l-2n}}{3+2l-2n}\left(\bar{z}-\bar{z}_k\right)^{n-l-2}\right.\nonumber\\
    &\phantom{=}\;\left.+\frac{2r_k^{3+2l}}{3+2l}\left(\bar{z}-\bar{z}_k\right)^{-n-l-2}\right]+\cc \ .\label{z_Tj}
\end{align}
We can also expand it in terms of $\bar{z}-\bar{z}_j$:
\begin{align}
    T_j&=-\mathrm{i}\oint_{C_j}\mathrm{d}z\int\mathrm{d}\bar{z}\sqrt{\frac{z-z_j}{\bar{z}-\bar{z}_j}}\nonumber\\
    &\phantom{=}\;\sum_{q=1}^3\sum_{m=0}^q\left[\frac{1}{(z-z_j)^m(\bar{z}-\bar{z}_j)^{q-m}}\sum_{l=0}^\infty C_{m,q,l}(\bar{z}-\bar{z}_j)^l\right]\nonumber\\
    &\phantom{=}\;-\mathrm{i}\sum_{k\neq j}\oint_{C_k}\mathrm{d}z\int\mathrm{d}\bar{z}\sqrt{\frac{z-z_k}{\bar{z}-\bar{z}_k}}\nonumber\\
    &\phantom{=}\;\sum_{n=0}^2\left[\left(\frac{1}{(z-z_k)^n}+\frac{1}{(\bar{z}-\bar{z}_k)^n}\right)\sum_{l=0}^\infty D_{n,l}(\bar{z}-\bar{z}_k)^l\right]+\cc\nonumber\\
    &=-\mathrm{i}\oint_{C_j}\mathrm{d}z\sum_{q=1}^3\sum_{m=0}^q\sum_{l=0}^\infty \frac{2C_{m,q,l}r_j^{1-2q+2m+2l}}{1-2q+2m+2l}\left(z-z_j\right)^{q-2m-l}\nonumber\\
    &\phantom{=}\;-\mathrm{i}\sum_{k\neq j}\oint_{C_k}\mathrm{d}z\sum_{n=0}^2\sum_{l=0}^\infty D_{n,l}\left[\frac{2r_k^{2l+1}}{2l+1}\left(z-z_k\right)^{-n-l}\right.\nonumber\\
    &\phantom{=}\;\left.+\frac{2r_k^{1+2l-2n}}{1+2l-2n}\left(z-z_k\right)^{n-l}\right]+\cc\label{zc_Tj} \ ,
\end{align}
which represents the order change of the area integral. $A$, $B$, $C$ and $D$ are the expansion coefficients, which can be obtained by expanding the integrand near the boundaries around each defects. And we should notice that
\begin{align}
    A_{m,q,0}=C_{m,q,0},\quad B_{n,0}=D_{n,0},
\end{align}
This relationship means Eqs. (\ref{z_Tj}) and (\ref{zc_Tj}) should give the same result, which reflects the order free of the integral with respect to $z$ and $\bar{z}$. It is easy to check that the leading term that contributes to $T_j$ is $-2A_{2,3,0}/r_j$ or $-2C_{2,3,0}/r_j$.

Finally, the expression of $T_j$ is
\begin{align}
    T_j=&\mathfrak{Re}\left[\frac{\pi q_0\alpha}{16R\zeta r_j}\exp{\left(2\mathrm{i}\Omega\right)}\left(4R^2+|z_j|^2\right)\sqrt{\frac{\bar{z}_j}{z_j}}\prod_{l\neq j}\sqrt{\frac{z_j-z_l}{\bar{z}_j-\bar{z}_l}}\right]\nonumber\\
    =&\frac{\pi q_0\alpha R}{4\zeta r}\cos \left(w_j+2\Omega-\phi_j\right).\label{cT_j}
\end{align}
We can perform the same procedure to the calculation of $S_j$ and $L$:
\begin{align}
    S_j=&\mathfrak{Re}\left[-\mathrm{i}\frac{\pi q_0\alpha\bar{z}_j}{4\zeta r_j}\exp{\left(2\mathrm{i}\Omega\right)}\prod_{l\neq j}\sqrt{\frac{z_j-z_l}{\bar{z}_j-\bar{z}_l}}\right]\nonumber\\
    =&\frac{\pi q_0\alpha R}{4\zeta r}\sin\theta_j\sin \left(w_j+2\Omega-\phi_j\right) \ ,\label{cS_j}\\
    L=&0 \ .
\end{align}
And $w_j$ is the summation of the phase angle between the $j$-th defect and others:
\begin{align}
    w_j=\sum_{l\neq j}\arg \left(z_j-z_l\right).
\end{align}

\bibliography{reference.bib}

\end{document}